# *Pbnm* to $R\bar{3}c$ phase transformation in *(1-x)*LaFeO$_3$.*x*LaMnO$_3$ solid solution due to modifications in structure, octahedral tilt and valence states of Fe/Mn

E. G. Rini[1], Mayanak. K. Gupta[2], R. Mittal[2,3], A. Mekki[4], Mohammed H. Al Saeed[4], Somaditya Sen[1*]

[1]*Department of Physics, Indian Institute of Technology Indore, Indore, 453552, India*

[2]*Solid State Physics Division Bhabha Atomic Research Centre Mumbai 400085, India*

[3]*Homi Bhabha National Institute, Anushaktinagar, Mumbai, 400094, India*

[4]*Department of Physics, King Fahd University of Petroleum & Minerals Dhahran, 31261, Saudi Arabia*

*\* Corresponding author: sens@iiti.ac.in (SS)*

**ABSTRACT:**

A theoretically supported experimental study of the *(1-x)*LaFeO$_3$.*x*LaMnO$_3$ (LFO-LMO) solid solution is being reported for the first time which reveals a phase transformation from the *Pbnm* and $R\bar{3}c$ phase at a chemical composition of *x*=0.625. Correlation of octahedral distortion and phase transition was extensively investigated using x-ray photoelectron spectroscopy (XPS), Raman and x-ray diffraction (XRD) measurements and density functional theory (DFT) calculation. A detailed study of the structural lattice parameters, bond lengths, bond angles have been done, supported by valence state and electronic properties studies. All the above parameters show a correlated modification to the phase transition. The distortion and tilting of the BO$_6$ octahedra has been studied as a function of different Fe:Mn content and expressed by Glazer representation from the refined Crystallographic Information Files (CIF). The angle of tilting from the central non-tilted position also shows a correlated modification with the phase transformation. The valence state and size of cations influences the octahedral tilting. Octahedral volume is reduced as the entire perovskite structure is relatively flattened with increasing Mn-content implying a flattening of both the BO$_6$ octahedra and the La$_8$O$_6$ cage. The vibrational properties were studied experimentally and supported by DFT phonon calculations, detailing the displacement pattern (eigen vectors) revealing considerable insight into the lattice dynamics of the compounds. The optoelectronic modifications in the band properties were studied experimentally and supported with theory. Hence, this manuscript is a in-depth analysis of the structure correlated phase transition of the LFO-LMO solid solution.

## I. INTRODUCTION:

ABO$_3$ materials demonstrate structural phase transitions which have generated a lot of interest in science and technology. Various physical phenomena such as ferromagnetism, ferroelectricity, multiferroicity, magnetoresistance are associated with ABO$_3$ perovskites [1-7] which exhibit an extraordinary range of structures, physical properties, and chemical bonding [8, 9]. There are various types of solid-state structural transitions [10, 11] depending on the choice of A and B cations [12, 13]. A lattice is cubic for a tolerance factor, $t = 1$ [14, 15] with B-O-B bond angle, $\varphi = 180°$ [16]. For $t < 1$, the cubic structure becomes unstable [17] due to increase in internal stresses [18] and the octahedra tends to tilt and rotate, leading to first a rhombohedral structure for $t \to 1$ [19] and further to an orthorhombic structure for very low $t$ ~ 0.8 [20]. It is observed that changes in B-O-B bond angles and hence, the rotations of the BO6 octahedra are correlated to elongated B-O bonds associated with compressed A-O bonds [21]. With a misfitted cationic change, $t$ deviates from unity. This is associated with the rotation of the BO6 octahedra about the [1 1 0] axis [22].

In this study, the intention is to analyze a continuous modification of structure from orthorhombic (*Pbnm*) LaFeO$_3$ [23] (hereafter called LFO) to rhombohedral ($R\bar{3}c$) LaMnO$_3$ [24] (hereafter called LMO), and correlate such changes to other physical parameters. Structure and other physical properties like electronic and magnetic properties are influenced by the ionic size and charge states of the constituent ions [25, 26]. These parameters may also participate in the octahedral rotation along the [1 1 1] axis [27].

The rotations, tilt and distortions of the BO$_6$ octahedra are modified by the changes in interactions between electronic orbitals [8, 28]. These factors are responsible for the modification of the crystal structure [8, 28], thereby bringing in changes in the density of electronic states. Hence, modifications of the A or B sites, involves new electronic hybridization [29, 30] and is therefore the source of exciting new electronic and magnetic properties leading to new applications [31, 32]. Therefore, the influence of each chemical modification and its correlation with the structure and properties is of prime interest. According to crystal field theory (CFT), electron orbital state degeneracies break down due to transition metal - oxygen interactions in perovskite oxides [33]. Hence, in this work an attempt has been made to understand the effect of combination of Fe and Mn at the B-site of the *(1-x)*LaFeO$_3$.*x*LaMnO$_3$ (LFO-LMO) solid solution (hereafter denoted as *LaFe$_{1-x}$Mn$_x$O$_3$* [LFMO]) from experimental aspects and substantiate the same from theoretical studies. The effect of ionic charge, ionic radii on octahedral distortion is discussed relating these to variations in structure, valence state, optoelectronic and vibrational properties of LFMO.

The end products LFO and LMO are extensively studied important materials from the aspect of applicability [6, 34-37]. Hence, a solid solution of these two in various proportions can open up new possibilities. Perovskite LFO is widely used as fuel cell cathodes [34], battery electrode materials [38], humidity sensors and alcohol sensors [39] and catalyst material [35]. A distorted perovskite LFO shows structural phase transitions at different temperatures. It has an orthorhombic structure (space group *Pbnm*) at room temperature. At ~1200 K, the structure becomes rhombohedral (space group $R\bar{3}c$) [40]. LFO is multiferroic [41], and possesses properties like exchange bias [42]. Hence, it is used in memory devices and spintronic devices. It is also used as a magneto-optic material [43], etc.. However, its unique structural properties are largely influenced by processing conditions [44]. On the other hand, perovskite rhombohedral ($R\bar{3}c$) LMO is a strongly correlated system [45] wherein the correlation among electrons can result in a variety of interesting properties [45, 46]. It finds applicability in spintronic [47], solid state fuel cell [36], magnetic sensors [37], etc.

A theoretically supported experimental study of the LFMO solid solution has not been studied yet. Using DFT phonon calculations, the displacement pattern (eigen vectors) provides considerable insight into the lattice dynamics of the compounds. A strong correlation between structural parameters like bond lengths, bond angles etc. with vibrational and optoelectronic properties has been proposed. Octahedral distortion and tilting as a result of different Fe:Mn content in the solid solution leads to a detailed structural study which has been substantiated with the tilting pattern of subsequent BO$_6$ octahedra as specified by Glazer representation for both the *Pbnm* and $R\bar{3}c$ phases.

## II. METHODOLOGY

### A. EXPERIMENTAL METHODS

Nanocrystalline LFMO compositions of chemical formula $LaFe_xMn_{1-x}O_3$ (with x=0.0, 0.125, 0.25, 0.375, 0.5, 0.625, 0.75, 0.875, 1.0, hereafter called LFO, LFMO12, LFMO25, LFMO37, LFMO50, LFMO62, LFMO75, LFMO87 and LMO, respectively), were synthesized using standard Pechini sol-gel method [48]. High purity precursors of La (Lanthanum (III) Oxide, Alfa Aesar, 99.9%), Fe (Iron (III) Nitrate Nona-hydrate, 98.0%) and Mn (Manganese (II) Carbonate, Alfa Aesar, 99.9%) were used for the synthesis of these materials. Iron nitrate was dissolved in deionized water (DIW), while $La_2O_3$ was dissolved in $HNO_3$ and DIW. $MnCO_3$ was dissolved in $HNO_3$ and DIW. In a separate beaker citric acid and glycerol were mixed to form a polymeric solution to be used for binding the precursor ions and also as a fuel. The solutions were poured one to the other to form polymeric mixtures of the desired compositions. These mixture solutions were stirred continuously while being heated at ~80°C to evaporate the water content. The sols were dehydrated to form gels. The gels were burnt in air. Dark brown or black powders were formed. These powders contained remnants of the nitrogenous and carbon compounds. To get rid of these unwanted elements the powders were heated in air at 450°C for 6 h. At this temperature, the desired phase was not obtained. The desired phase was ultimately formed by further heating the powders, first at 600°C and thereafter at 750°C.

A Bruker D2 Phaser Powder X-ray Diffractometer equipped with a Cu $K_\alpha$ source ($\lambda = 1.54$ Å) and a Charge-Coupled-Device(CCD) detector was used to study the structural properties of LFMO for $2\theta \sim 20°$ to $80°$, i.e. d-spacing ~ 4.4A to 1.1A. Structural properties, lattice parameters, and density were extracted from Rietveld refinement using GSAS software with orthorhombic *Pbnm* and rhombohedral $R\bar{3}c$ space groups.

Raman spectra for LFMO samples were measured at room temperature. A Jobin-Yvon Horiba LABRAM-HR Micro-Raman spectrometer was used. A red light excitation of wavelength ~ 622 nm was obtained from a laser source.

The chemical composition of the LFMO samples and the oxidation states of the elements were evaluated from X-ray Photoelectron Spectroscopy (XPS) using a Thermo-Scientific Escalab 250 Xi XPS Spectrometer. The energy resolution was ~ 0.5 eV. Monochromatic Al-$K\alpha$ x-rays were used to irradiate the LFMO samples. A flood gun was used to neutralize charging effects. The $Ar^+$ ion bombardment of the samples was performed in the XPS UHV chamber by applying acceleration potentials of 3 kV for 60 seconds. A survey scan was performed to assess all the elements present. The Mn 2$p$, O 1$s$, La 3$d$, Fe 2$p$ and C 1$s$ high resolution XPS spectra were acquired. The base pressure in the chamber was originally $1\times10^{-10}$ mbar. However, during the ion bombardment, the pressure was ~ $1\times10^{-6}$ mbar. The energy scale was calibrated using the binding energy of adventitious carbon (C 1s = 284.8 eV). The experiments were repeated in order to check for reproducibility of the results. "XPSFit" was used to fit the data.

The optical and bandgap properties have been analyzed using a UV-visible spectroscopy (Research India UV-Vis spectrometer) in the range 200 to 800 nm.

**B. THEORETICAL METHODS**

All the calculations were done using Quantum Espresso software suite [49]. We have used the projected augmented wave (PAW) [50] flavor of pseudo potential within generalized gradient approximation (GGA) [51] parameterization by Perdew, Burke and Ernzerh [52]. A kinetic energy cutoff of 550 eV for the plane wave pseudopotentials is used. The 6×6×4 k-point mesh generated using Monkhorst-Pack method [53] was used to integrate the Brillouin zone. All results converge satisfactorily with respect to k mesh and energy cutoff (550 eV) for the plane-wave expansion. (6x6x4) and (8x8x8) k-points mesh were generated according to the Monkhorst-Pack (MP) scheme for *Pbnm* and $R\overline{3}c$ respectively. A convergence criterion of 10-8 eV and 10-4 eVÅ -1 were chosen for total energy and ionic forces, respectively. The valence electron configurations of O, La, Fe, and Mn, as used in calculation for pseudopotentials generation are $s^2p^4$, $s^2p^6d^1s^2$, $p^6d^6s^2$ and $p^6d^5s^2$ respectively [48]. The phonon frequencies have been calculated using density functional perturbation theory (DFPT) [54].

**III. RESULTS AND DISCUSSIONS**

   **A. Structural studies:**

The XRD data of the samples reveal multiple peaks which can confirm the structural similarities of these samples to that of either orthorhombic *Pbnm* LFO or rhombohedral $R\overline{3}c$ LMO [Fig. 1(a)]. For LFO, a reflection at 32.39° corresponding to an orthorhombic *Pbnm* [112] reflection is observed. This reflection shifts to a lower angle for *x*=0.125 [Fig. 1(b)]. Thereafter, the peak shifts to higher angles with increasing Mn-content, for 0.25≤*x*≤0.5. A very weak *Pbnm* [111] reflection is observed at 25.7° for 0≤*x*≤0.375. However, the reflection disappears for 0.5≤*x*≤1.0 [Fig. 1(c)]. All these indicate an orthorhombic *Pbnm* phase only in the regime 0≤*x*≤0.5, which is absent in compositions with higher Mn-content.

For x=0.625 the [112] peak splits into two corresponding to the [2,-1,0] and [1,0,4] reflections of the rhombohedral $R\bar{3}c$ LMO structure. Hence, this splitting indicates an orthorhombic to rhombohedral (*Pbnm* → $R\bar{3}c$) phase transition. The two peaks move towards higher angles for 0.625≤*x*≤1.0. Similar observations are obtained for other weaker peaks at higher angles [Fig. 1(d)].

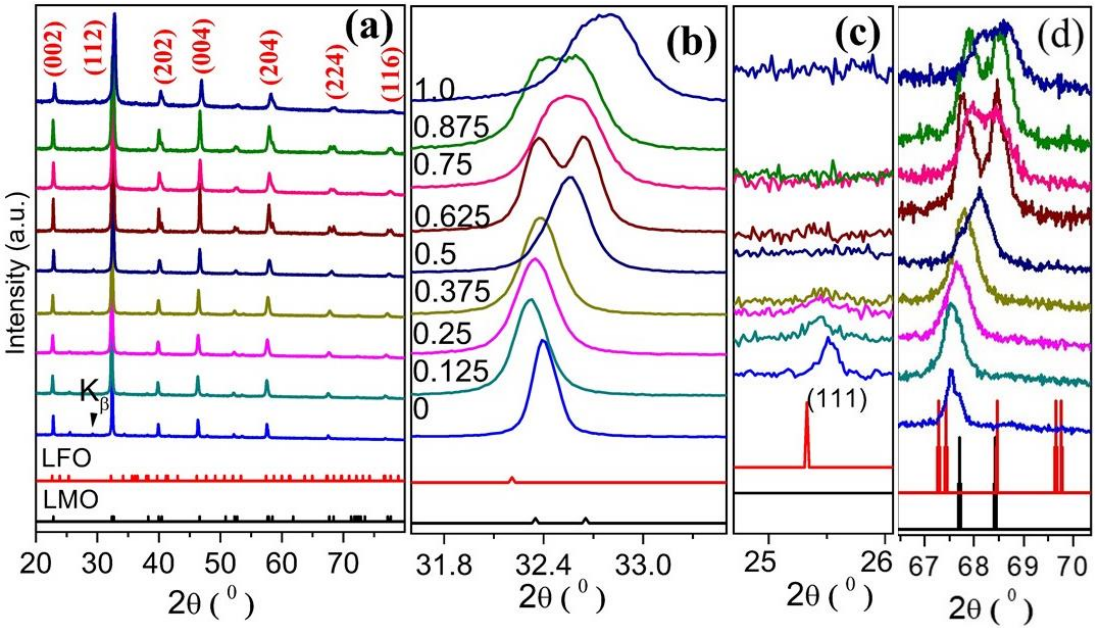

**Fig.1:** (a) X-ray diffraction (XRD) pattern reveals a pure phase for LFMO (0 ≤ *x* ≤ 1.0) (b) Shifting of LFO (112) peak and splitting observed for 0.625≤*x*≤1.0 (c) weak (111) peak observed for 0≤*x*≤0.375 (d) Splitting of (224) peak observed for 0.625≤*x*≤1.0.

Rietveld refinement confirms a *Pbnm* → $R\bar{3}c$ phase transition between *x*=0.5 and *x*=0.625. The refined structural parameters are shown in Table S1(a and b). The refinement also provides information on the thermal parameters of each structural site. The simulated data fits well with the experimental data for all samples, indicating the absence of impurity secondary phases [Fig. S1]. Reasonably good fitting parameters, $R_p$ and $R_{wp}$ were obtained ~ <5%. However, only for *x*=0.875 these values were >5%; $R_p$=4.9%/$R_{wp}$=6.53%. From the refinement, the density of the LFMO samples was observed to increase with increasing Mn content in the *Pbnm* phase but remained almost constant in the $R\bar{3}c$ phase.

For the *Pbnm* phase, all three lattice parameters ($a_O$, $b_O$, $c_O$) decreased with increasing Mn-content [Fig. 2(a)]. For the $R\bar{3}c$ phase, $c_R$ increases from 13.36A (in LFMO62) to 13.396A (in LMO). However, $a_R$ and $b_R$ continues to decrease with increasing Mn-content. To be noted, the rate of decrease of $a_O$ and $b_O$ axes is higher than $a_R$ and $b_R$. A sudden increase of $c_R$ with a continual decrease in $a_R$ and $b_R$ is an indication of the increase of rhombohedral distortion along the *c*-axis.

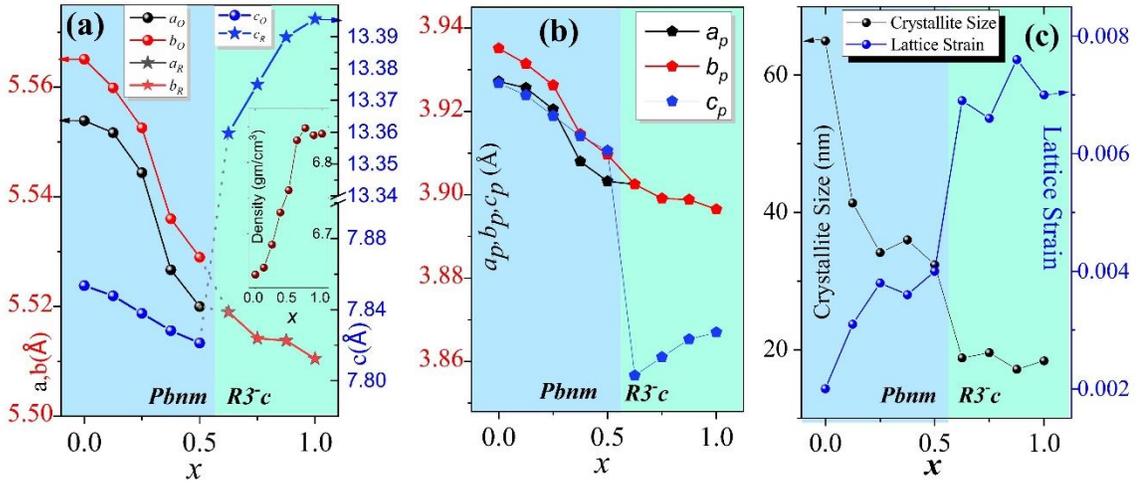

**Fig. 2:** (a) Lattice parameters $a_O$, $b_O$, $c_O$ of the orthorhombic *Pbnm* phase (0≤*x*≤0.50) decreases with increasing Mn content and for the $R\bar{3}c$ phase (0.625≤*x*≤1.0), $a_R$ and $b_R$ continues to decrease whereas, $c_R$ increases with Mn content revealing a dependence on the B-site ionic radius; (b) the pseudocubic lattice parameters, $a_p$, $b_p$ and $c_p$ for the orthorhombic (*Pbnm*) phase and rhombohedral ($R\bar{3}c$) lattice parameters, $a_R$, $b_R$, $c_R$; (c) increment of lattice strain correlated with decrement of crystallite size with B-site substitution.

However, one must recognize that $a_O$, $b_O$, $c_O$ are not directly comparable to $a_R$, $b_R$ and $c_R$. To understand an equivalent structure of both the phases one must look into the pseudocubic structures. The orthorhombic *Pbnm* axes [100] and [010] coincide with the [110] and [1$\bar{1}$0] pseudo cubic axes, and the [001] is parallel to the [001] pseudocubic axis [55]. The other axes are reoriented [Fig. 3(a)] For the rhombohedral $R\bar{3}c$ phase the [110], [011] and [101] axes coincide with that of the pseudo cubic axis [56] while the other axes are modified [Fig. 3(b)]. The lattice parameters, $a_O$, $b_O$, $c_O$ and $a_R$, $b_R$, $c_R$ can be expressed in terms of pseudocubic lattice parameters $a_p$, $b_p$ and $c_p$ as: $a_O \sim \sqrt{2}a_p$, $b_O \sim \sqrt{2}b_p$, $c_O \sim 2\,c_p$ and $a_R \sim \sqrt{2}a_P$, $b_R \sim \sqrt{2}b_P$, $c_R \sim 2\sqrt{3}c_P$ respectively [57]. The variations of $a_p$, $b_p$ and $c_p$ for $0 < x < 1.0$ were analysed for both the phases [Fig. 2(b)].

The pseudocubic lattice parameters, $a_p$ and $b_p$ decrease with increase in Mn content for the entire range $0 \leq x \leq 1.0$. For $0 \leq x < 0.625$, the pseudocubic lattice parameter, $a_p < b_p$, whereas for $x \geq 0.625$, the pseudocubic lattice parameter, $a_p = b_p$. However, $c_p$ decreases for $0 \leq x < 0.625$. At the phase transition, in between x=0.5 to 0.625, $c_p$ decreases drastically from 3.911(A) ($x$=0.50) to 3.857 (A) ($x$=0.625) and thereafter increases continuously in the range $0.75 \leq x \leq 1.0$ to 3.8669 (A) for $x$=1. Note that $c_p \sim a_p$ for $x \leq 0.25$ and $\sim b_p$ for $0.375 \leq x \leq 0.50$. In the rhombohedral phase that $c_p$ is shorter than both $a_p$ and $b_p$ (i.e. $c_p < a_p, b_p$). Hence, the pseudocubic structure keeps on shrinking and flattening for the entire range $0 \leq x \leq 1.0$, indicating a volume reduction.

Lattice strain was estimated from XRD data using Williamson–Hall equation [58]. This was done with the help of EVA software [48]. Lattice strain increases with increasing Mn content [Fig. 2(c)]. The crystallite size was also estimated using the Scherrer equation [59, 60]. With increasing Mn-content the crystallite size decreases from 65nm in LFO to 18.5nm in LMO [Fig. 2(c)]. The decrease of crystallite size may be due to the increase in strain in the lattice.

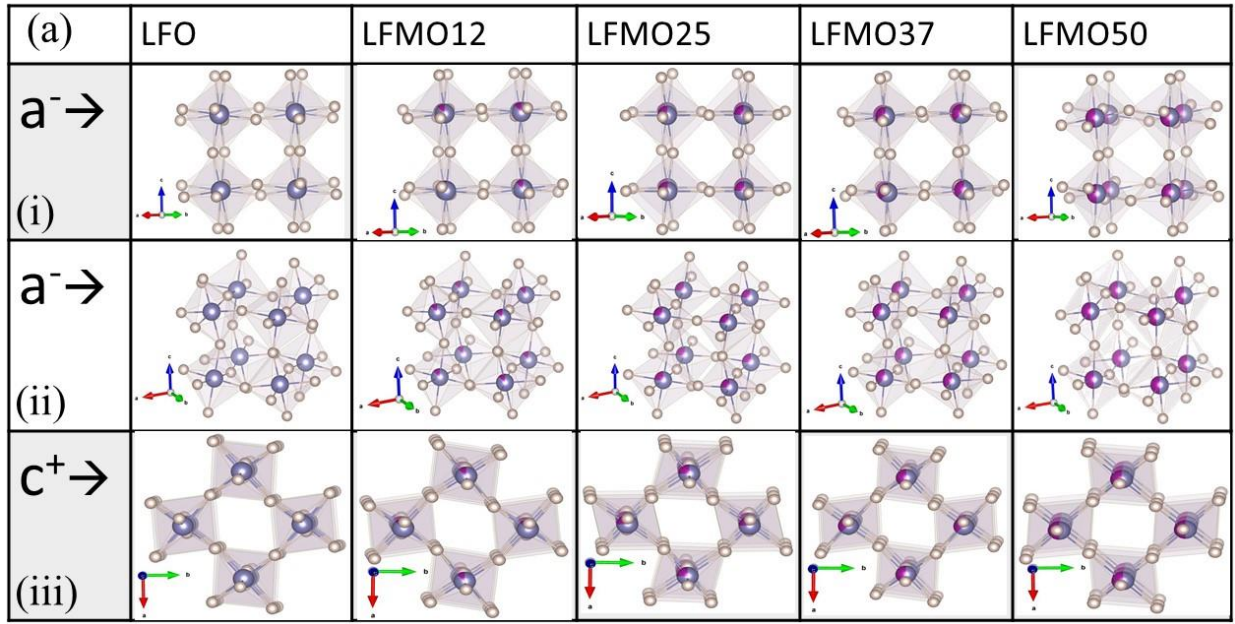

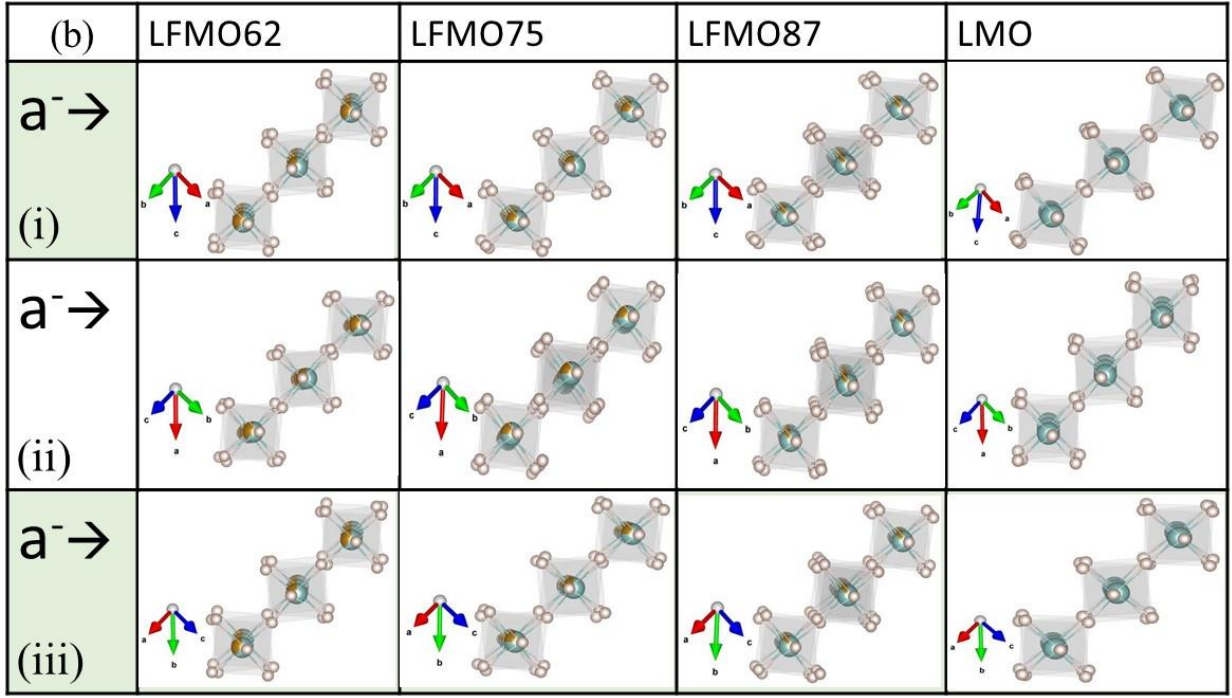

**Fig. 3**: (a) Glazer representation ($a^-a^-c^+$) observed for *Pbnm* phase for $0 \leq x \leq 0.50$, and (b) ($a^-a^-a^-$) observed for $R\bar{3}c$ phase for $0.625 \leq x \leq 1.0$.

The $BO_6$ octahedra are tilted in different directions in these two phases. Hence, these tilting differences in terms of rotation and sequence of tilting can only be evaluated by understanding the pseudocubic structures in greater detail and thereby finding the Glazer notations of the same [61]. By viewing the structure along the orthorhombic *Pbnm* [110] and [1$\bar{1}$0] axes, one can see antiphase tilting of successive octahedra down these axes [62] [Fig. 3(a) i and ii]. However, along the orthorhombic *Pbnm* [001] axis, an in-phase tilting of successive octahedra down this axis can be viewed [Fig. 3(a) iii]. Hence, the Glazer representation is ($a^-a^-c^+$) for the *Pbnm* phase (for $0 \leq x \leq 0.50$) [Fig. 3 (a)] in agreement with previous reports [55] meaning tilts of equal magnitude but in opposite direction for subsequent octahedra along the pseudo cubic *a* and *b* axes. Due to the equal tilts, the letter *a* has been repeated in the first and second position.

By viewing the structure along the rhombohedral $R\bar{3}c$ [110], [011] and [101] [56] axes (for $0.625 \leq x \leq 1.0$), one can see equal antiphase tilting of successive octahedra down these axes [62] [Fig. 3(b) i, ii and iii]. Hence, for the rhombohedral $R\bar{3}c$ phase [Fig. 3 (b)] (for 0.625

≤ x ≤ 1.0) the Glazer notation is ($a^-a^-a^-$), consistent with earlier reports rhombohedral $R\bar{3}c$ distorted perovskites [56].

To understand these changes more thoroughly, a detailed investigation was performed involving the bond lengths and angles involved in between the ions including the cations and anions. To achieve the above, the refined crystallographic index files were visualized using Mercury software. However, before going into the analysis of the bond lengths and bond angles it is beneficial to have a clear estimation of the ionic valence states of the constituent ions, especially the B-site ions. The ionic size of the ions will be modified according to the valence state of the ions. The hybridization strength will also be affected by the valence state and hence will affect the bond lengths and hence the bond angles.

### B. Valence States Studies

The high spin six-coordinated $Fe^{3+}$ and $Mn^{3+}$ ions have a similar ionic radius of 0.785A. However, the low spin six-coordinated $Mn^{3+}$ ion is larger ~0.72A than the low spin $Fe^{3+}$ ion ~0.69A, while $Mn^{4+}$ is much smaller ~0.67A compared to all forms of $Fe^{3+}$ and $Mn^{3+}$. Therefore, a high spin (hs) state of $Fe^{3+}$ in LFO, followed by any sort of Mn-substitution should lead to lattice contraction and thereby introduce stress in the lattice. However, if $Fe^{3+}$ is in the low spin (ls) state then only $Mn^{4+}$ state can be held responsible for a lattice contraction as well as increased strain. To understand the valence states and the spin arrangements of the individual elements a proper XPS analysis has been performed on the samples. The XPS survey scan [Fig. 5(g)] confirms the presence of only La, Fe, Mn, and O in the LFMO samples in agreement with the elemental mapping analysis. No other elements are observed apart from these constituent elements confirming the purity of these samples in terms of chemical composition.

#### i. Mn-2$p$ core level spectra: results

The Mn-2$p$ core level XPS spectra [Fig. 4(b)] reveals two multi-component features [Fig. 4(e)] at ~ 642 eV and ~654 eV. After a Tougart-type background correction [63] each feature can be deconvoluted into two mixed Gaussian-Lorentzian shaped peaks: (642.3eV, 644.02 eV) and (653.52eV, 656.97eV). A Mn-2$p_{3/2}$ feature centered at binding energy ~642 eV and Mn-2$p_{1/2}$ at ~ 653.5 eV is an indicator of the presence of $Mn^{3+}$ ions [64] with a spin-orbit split equal to ~11.7eV [65], while at binding energy ~642.4eV and 654.1 eV is an indicator of the presence of $Mn^{4+}$ ions [64] with a spin-orbit split equal to ~11.7 eV [66].

Assuming a similar logic, in these samples, the presence of the Mn-$2p_{3/2}$ and Mn-$2p_{1/2}$ contributions at ~ 642.3 eV and 653.52 eV with a difference of ~11.2 eV represents the presence of $Mn^{3+}$ states, while the presence of the Mn-$2p_{3/2}$ and Mn-$2p_{1/2}$ contributions at ~644.02 eV and 656.97 eV with a difference of ~12.9 eV represents the presence of $Mn^{4+}$ states, for all samples. Hence, Mn ions were in a mixed valence state of $Mn^{3+}$ and $Mn^{4+}$ states for all samples [Fig. 4(b)]. There is a net increasing trend of the area of the peaks belonging to the $Mn^{4+}$ ions with respect to that of $Mn^{3+}$ with increasing Mn content. There are certain fluctuations in the trend, but overall an incremental nature is observed [Fig. S2.i(a)]. The $Mn^{3+}$ contribution reduces from ~90% in $x$=0.125, to ~80% in $x$=1 sample, while $Mn^{4+}$ contribution increases from ~10% in $x$=0.125, to ~20% in $x$=1 sample. Hence, as the proportion of quadrivalent $Mn^{4+}$ to trivalent $Mn^{3+}$ and $Fe^{3+}$ increases, the O-content should increase in the lattice. The Mn2$p$ XPS spectra reveals a decrease of $Mn^{3+}$ and increase of $Mn^{4+}$ with increasing Mn until $x$=0.625 and thereafter almost remains constant. At $x$=0.625, the structure changes to rhombohedral ($R3^-c$) to accommodate more oxygen due to extra charge on Mn (i.e. $Mn^{4+}$). Beyond $x$=0.625, the rhombohedral structure doesn't allow oxygen to further accommodate. The ratio (3/2)/(1/2) reaches an ideal value of 2 at $x$~0.5 [Fig. S2.i(b)]. At $x$=0.625, $Mn^{4+}$ becomes maximum. $Mn^{4+}$ contains 1 electron less as compared to $Mn^{3+}$ in the outermost orbit and hence this ratio increases.

Satellite peaks are observed at 647.01 eV and 648.76 eV. Satellite peaks are reported for $Mn_2O_3$ at (~646.2 eV) for $Mn^{3+}$ state [67, 68] and $MnO_2$ at (~646.4 eV) for $Mn^{4+}$ state [68, 69]. Satellite peaks are observed when photoelectron interact with outermost 3$d$ electron (closely packed multiplet split-3$d$ electron). As a result the outermost 3$d$ electron after gaining energy from photoelectron get excited to higher vacant state [70]. The Mn2$p$ satellite intensity decreases for 0.125≤$x$≤0.5, and thereafter increases for 0.5≤$x$≤0.875. However, for $x$=1.0, the intensity decreases again. The $Mn^{3+}$ satellite% decreases to a minimum for x=0.5 [Fig. S2.i(d)]. As structure becomes rhombohedral at x=0.625, Mn-O bonds change and hence satellite% increases for $x$=0.625. For further increase in x in the rhombohedral phase, the satellite% decreases for 0.625≤x≤1.0. Hence, Mn-incorporation in the LFO lattice reduces multiplet splitting of the Mn3$d$ electrons. The splitting is minimum at $x$=0.5 for the orthorhombic phase. There is a sudden increase of splitting at the orthorhombic to rhombohedral phase transition, i.e. for x=0.625. In the rhombohedral phase a similar decreasing splitting effect is seen with increasing Mn-content. Hence, the phase transition demonstrates an increased multiplet splitting of the Mn3*d* electrons, but generally with increase of Mn-content the Mn3*d* electrons loose the splitting tendency.

**ii.Fe-2*p* core level spectra: results**

The Fe-2*p* core level XPS spectra [Fig. 4(a)] reveals two multi-component features [Fig. 4(d)] at ~712 eV and ~725 eV. These peaks are typically observed for $Fe_2O_3$ [71]. A Fe-2$p_{3/2}$ feature centered at binding energy ~ 711 eV [71] and Fe-2$p_{1/2}$ at ~ 724.6 eV [71] is an indicator of the presence of $Fe^{3+}$ ions with a spin-orbit split equal to ~13.5eV [72, 73]. Assuming a similar logic, in these samples, the difference between the Fe-2$p_{3/2}$ at ~ 711.78 eV and Fe-2$p_{1/2}$ at ~ 725.34 eV contributions was ~13.5 eV for all samples. Hence, Fe ions were predominantly in the $Fe^{3+}$ state for all samples [Fig. 4(a)]. A Tougart-type background correction was done for all the samples [63]. The total Fe-2*p* features could be further deconvoluted into five Gaussian peaks [Fig. 4(d)] due to the presence of Fe-2$p_{3/2}$, Fe-2$p_{1/2}$ and a satellite feature [71, 74]. The literature reveals that Fe-2$p_{3/2}$ and Fe-2$p_{1/2}$ peaks for low-spin iron (II) are narrow and are separated by -13 eV, whereas for high-spin iron (III) the features are broader, separated by -13.5 eV and shows an extensive satellite structure. The broadness of the spin components is due to multiplet splitting and is a cause for the satellite feature [75, 76].

Satellite peaks of Mn and Fe states:

With increase in Mn content, the satellite % due to interaction of Fe-2*p* photoelectron with outermost 3*d* electron decreases for $x \leq 0.5$. At $x=0.625$, due to rhombohedral structure Fe-O bond strength changes and hence satellite% increases. Further, the ratio ($2p_{3/2}/2p_{1/2}$) reaches an ideal value at $x \sim 0.5$ [Fig. S2.i(c)] as observed in Mn-2*p* XPS analysis. The origin of satellite features is due to various factors like shakeup peak, coupling between the partly filled 3d shell and the ejected photoelectron etc. [76]. The core 2*p* photoelectron during their transit to continuum state transfers a part of kinetic energy to 3*d* electron (valence band electron) and valence electron after gaining this energy transit from this band to empty 4s state. This energy difference is observed while core electron transit to continuum state on passing through valence band is seen in shake-up state. However, the Fe-2*p* features are prominent for only $x < 0.375$. Beyond $x = 0.375$, the features become extremely weak and become undetectable $x > 0.5$ and analysis becomes difficult. The satellite contribution from the Fe edge is highly modified by the Mn-content and it seems that both Fe and Mn states try to influence one another [Fig. S2.i(e)].

**Correlation of XPS analysis with XRD analysis and probable spin state of Fe and Mn:**

From the average B-O bond length, the B-site radius was observed to decrease with increase in Mn-content in the orthorhombic phase. In the rhombohedral phase the B-site radius starts to increase nominally in the regime $0.625 \leq x \leq 0.875$, and thereafter reduces suddenly. From XPS analysis, the $Mn^{4+}$ content becomes invariant at ~19.9% in the rhombohedral phase for x>0.50. However, this is quite a low $Mn^{4+}$ percentage to support the experimentally observed ionic radius value in the rhombohedral phase as shown in the graph (green solid star) [Fig. S2.ii]. Only increased $Mn^{4+}$ proportions can somewhat give more logical values. An attempt has been made to evaluate the theoretical content of $Mn^{4+}$ with different combinations of $Fe^{3+}$ and $Mn^{3+}$ spin states that can replicate the ionic size obtained from XRD analysis. However, to match the experimental data the proportion of $Mn^{4+}$ had to be increased to approximately 56% in LMO. Hence, such high values of $Mn^{4+}$ may not be logically possible to justify since $Mn^{4+}$ will invite extra O into the lattice.

The B-site effective ionic radius was also calculated considering four possible combination with Fe and Mn spin states along with $Mn^{4+}$ contribution: ($Mn^{3+}hs + Fe^{3+}hs$), ($Mn^{3+}ls + Fe^{3+}hs$), ($Mn^{3+}hs + Fe^{3+}ls$), ($Mn^{3+}ls + Fe^{3+}ls$) and ($Mn^{4+} + Fe^{3+}hs$). On the basis of this analysis of the effective ionic radius, it may be mathematically derived that most probably, in the orthorhombic phase ($0 \leq x \leq 0.5$) Fe is mostly in the low spin state (~50-60%) while in the rhombohedral phase ($0.625 \leq x \leq 1.0$) it is in the low spin state (100%). Mn on the other hand is mostly in the low spin state.

### iii. O$1s$ core level spectra

The O$1s$ XPS spectra shows two features, commonly discussed in literature to be belonging to two chemical states: ~529.41 eV (lattice oxygen $O_{lat}$) and 531.42 eV (adsorbed oxygen $O_{ad}$) [Fig. S2.i(f) ] [77, 78]. A Tougart-type background [63] was subtracted. The background subtracted data can be deconvoluted into two asymmetric Gaussian wide peaks. The fitted data reveals that with increase in Mn content $O_{ad}$ increases while $O_{lat}$ decreases continuously. As a result, the $O_{ad}/O_{lat}$ ratio increases with Mn content [Fig. S2.i(g)]. The reason behind the increase of this ratio may be due to reduction of crystallite size which increases more surface area and hence more adsorption surface. Note that this increase of the $O_{ad}/O_{lat}$ ratio is not contradictory to the possibility of increase of $O_{lat}$ which should be also increasing due to increasing $Mn^{4+}$ contribution. Due to the fast decrease of crystallite size and thereby increase of $O_{ad}$ contribution this ratio seems to be increasing.

### iv. La-3d core level spectra

The La-3d core level XPS spectra reveals two multi-component features at ~836 eV and ~853 eV contributions (Fig. 4(c)). These peaks are typical of $La_2O_3$ [79]. A La-$3d_{5/2}$ feature centered at binding energy ~834.3 eV and La-$3d_{3/2}$ at ~851.0 eV is an indicator of the presence of $La^{3+}$ ions with a spin-orbit split equal to ~ 16.8 eV [80]. Assuming a similar logic, in these samples, the difference between the La-$3d_{5/2}$ at ~ 836 eV and La-$3d_{3/2}$ at ~ 853 eV contributions was ~16.8eV for all samples. Hence, La ions were predominantly in the $La^{3+}$ state for all samples [Fig. 4(c)]. A Tougart-type background correction was done for all the samples [63]. Each spin-orbit component (La-$3d_{3/2}$ and La-$3d_{5/2}$) could be further deconvoluted into three Gaussian peaks [Fig. 4(f)] due to the presence of multiplet states [81]. The La-$3d_{5/2}$ feature was fitted with three peaks centered at ~836 eV, 838 eV, 840 eV. The La-$3d_{3/2}$ feature was fitted with peaks centered at ~853, 855 eV and 857 eV. The first peak centered at 836 eV of La-$3d_{5/2}$ and 853 eV of La-$3d_{3/2}$ are the core level peaks and other two peaks centered at (838 eV and 840 eV) of La-$3d_{5/2}$ and (855 eV, 857 eV) of (La-$3d_{3/2}$) are the 'shakeup satellites' [81]. During the excitation of the 3d electron into the continuum state, this photoelectron transfers a part of the kinetic energy to the valence band electron while passing through this valence band. As a result of this interaction of $3d$ electron with the valence band electron, a part of kinetic energy is gained by valence state electron. These electrons after gaining energy transit from the valence band to empty 4f state [82, 83]. This is the difference of energy that is observed in satellite peak [82, 83]. Hence, the lower binding energy peak 836 eV and 853 eV of La-$3d_{5/2}$ and La-$3d_{3/2}$ is indicative of the presence of a core hole with the absence of electrons in the 4f orbital. A plasmon loss peak occurs at ~848.5 eV [81]. This is a consequence of the interaction between the photoelectron and other electrons which causes loss of a specific amount of energy [81, 84].

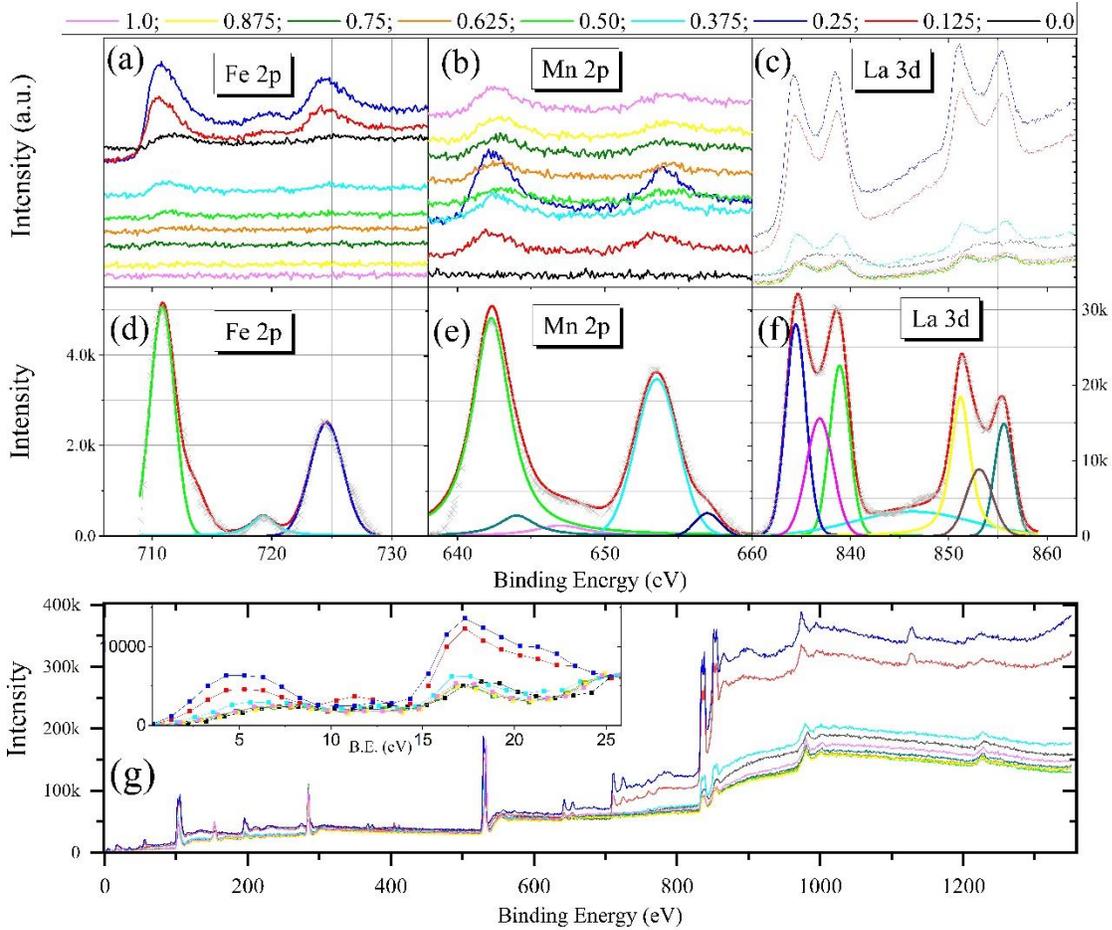

**Fig. 4**: XPS spectra of La(FeMn)O$_3$ nanoparticle. (a) La-3*d* (b) Fe-2*p* (c) Mn-2*p* (d) Deconvolution of La-3*d* (e) Deconvolution of Fe-2*p* (f) Deconvolution of Mn-2*p* (g) Survey

## C. Bond length and Bond angle Analysis

In this bond length/angle analysis the B-sites are invariant (symmetry constraints). Therefore, the Fe/Mn related bond lengths/angles will be discussed as a reference. As these are pseudocubic perovskite structures, a BO$_6$ octahedra is enclosed inside a La$_8$O$_6$ cuboidal cage [Fig. S3] for both the space groups. Each face of the La$_8$O$_6$ cuboid contains four La ions and a O ion located somewhere near the center of the face. It is observed that for the orthorhombic *Pbnm* phase, four out of the six sides of the cuboidal La$_8$O$_6$ cage are similar with similar bond lengths/angles. The remaining two are different from the previously mentioned four sides. However, these two sides are similar to each other in terms of bond lengths/angles. A line perpendicular to these two planes and through the B-site is considered a reference axis in this study. The reference axis is parallel to the pseudocubical c-axis. The oxygen ions in these two planes through which the reference axis passes will be called apical oxygen, $O_a$, while the other four are planar oxygens, $O_p$. None of the $O_p$-$O_p$-$O_p$, $O_p$-$O_a$-$O_p$ and $O_p$-B-$O_p$ bond angles are equal to 90° [Fig. 5(e)]. This ensures a distortion of the $BO_6$ octahedra.

It should be noted that in the rhombohedral $R\bar{3}c$ phase all six sides are equivalent, being constituted of comparable sets of bond lengths/angles. The octahedra seem to be less distorted with the $O_p$-$O_p$-$O_p$ and $O_p$-$O_a$-$O_p$ bond angles being equal to 90°. However, the $O_p$-B-$O_p$ angles in the $R\bar{3}c$ phase retain the non-90° nature, at par with the $Pbnm$ structure. Hence, although the O-ions seem to be better ordered, a distorted B-site is yet a feature of the $R\bar{3}c$ structure.

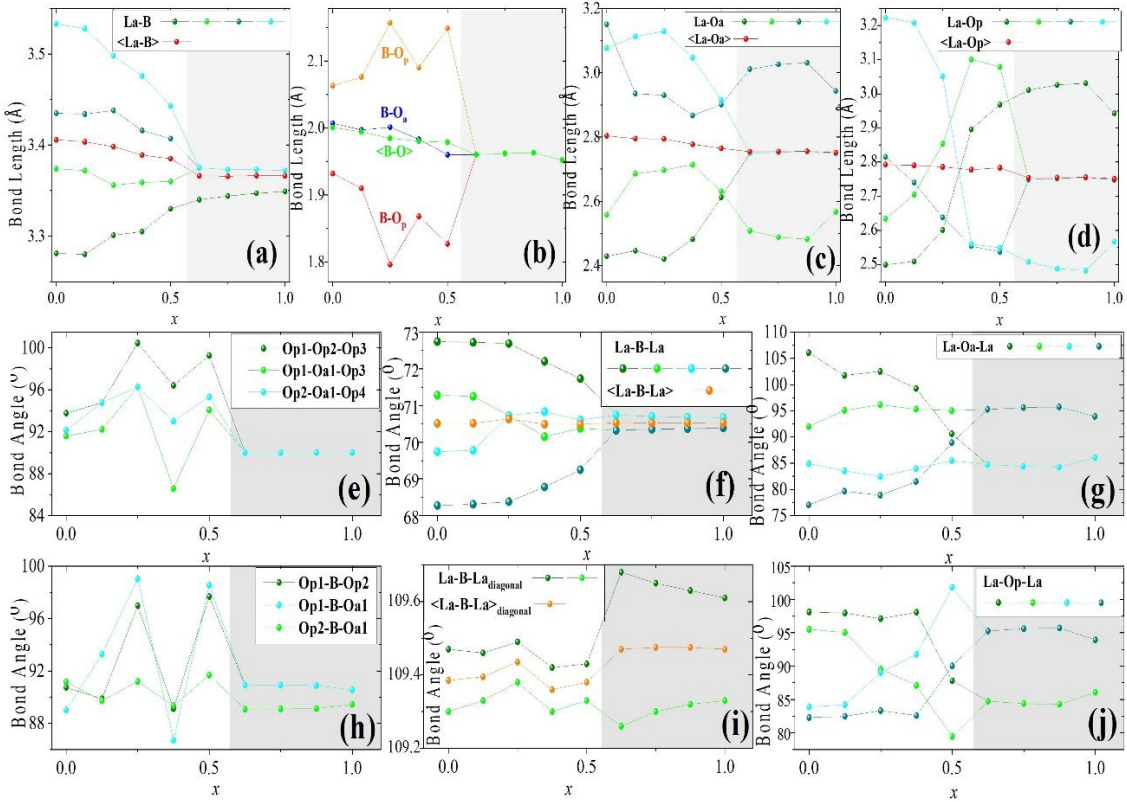

**Fig. 5**: (a) Convergence of the La-B bond lengths at $x = 0.5$ and the average $<$La-B$>$ bond length gradually decreases for $0 \leq x \leq 0.50$ and negligible changes are observed for $0.625 \leq x \leq 1.0$. (b) Continuous decrement of average bond length $<$B-O$>$ for $0 \leq x \leq 1.0$ whereas one B-Op bond increasing while the other one decreasing for $0 \leq x \leq 0.50$ and negligible changes are observed for $0.625 \leq x \leq 1.0$ and the B-$O_a$ bond being invariant for $0 \leq x \leq 1.0$. (c) Convergence and divergence observed for La-$O_a$ at $x = 0.5$ and average $<$La-$O_a$$>$ remains invariant with substitution (d) Variation of La-$O_p$ with Mn content (e) Increment and decrement observed for ∠$O_p$-$O_p$-$O_p$ ($x \leq 0.50$) and invariant in the range $0.625 \leq x \leq 1.0$. (f) A gradual convergence of the ∠La-B-La bond angles beyond $x = 0.5$ is observed. (g) Convergence and divergence observed for ∠La-B-La bond angles at $x = 0.5$ (h) Variation of ∠$O_p$-B-$O_p$ / ∠$O_p$-B-$O_a$ with Mn content. (i) Modification of diagonal ∠La-B-La bond angles with substitution (j) Variation of ∠La-$O_p$-La with Mn content.

The La-O bond length varies [Fig. 5(c and d)] with the strength of O$2p$-La$4f$ hybridization. On the other hand the O$2p$-Mn$3d$ and O$2p$-Fe$3d$ hybridization strengths modify the B-O bond distances. The resultant hybridization-forces on the O ion will modify the position of the ion w.r.t. the La and B ions. The distortion of the octahedron is linked to such forces and results in the non-90°-ness of the bond angles. A complete understanding of how these variations modify with the incorporation of Mn in LFO and ultimately lead to a *Pbnm* to $R\bar{3}c$ phase transition will be detailed below.

The B-$O_a$-B and the B-$O_p$-B angles show different trends in the *Pbnm* and $R\bar{3}c$ phases with increasing Mn-content [Fig. S4(b)]. These angles are related to the relative tilting of subsequent octahedra w.r.t each other along the c-axis (B-$O_a$-B angle) or along the a-b axes (B-$O_p$-B angle). The tilting angle is expressed as the difference of a B-O-B angle from 180°. While the B-$O_a$-B angle shows an increasing trend in the *Pbnm* phase, it reduces with reducing Fe content in the $R\bar{3}c$ phase. For pure LMO this angle is maximum. On the other hand, for x$\leq$ 0.25 the B-$O_p$-B angle increases but for x$\geq$0.25 decreases in the *Pbnm* phase. Thereafter, the B-$O_p$-B bond angle also decreases with decreasing Fe-content in $R\bar{3}c$ phase. As a result of these two different trends, the average B-O-B bond angle increases continuously in the *Pbnm* phase and thereafter reduces in the $R\bar{3}c$ phase. It is to be noted that in the $R\bar{3}c$ phase the B-$O_a$-B and the B-$O_p$-B angles are equal to one other, resulting in uniformity of all B-O-B bond angles. Although the relative tilting has individual trends along different directions, the average tilting which can be related to distortion has a definite trend of increase with substitution in the *Pbnm* phase followed by a trend of decrement in the $R\bar{3}c$ phase. This distortion may be a resultant of a strong hybridization due to overlap of B-$3d$ and O-$2p$ orbitals [85]. The distortion may vary depending on the ionic valency and spin. The average tilting angle decreases in the *Pbnm* phase and increases again in the $R\bar{3}c$ phase. For $x$=1, the tilting angle is minimum.

There are four distinct types of La-B bond lengths for the *Pbnm* phase with different values and relative angles. These four bonds modify with Mn incorporation and a tendency to converge is observed [Fig. 5(a)]. After a phase transition to the $R\bar{3}c$ phase, three of the larger bond lengths coincide while the shortest fourth increases gradually to nearly equate the other three. These changes may be correlated to the similar converging trends of the La-B-La bond angles [Fig. 5(f)]. This regularity in the pattern of changing interaction between the La and B ions can be correlated to the ionic properties of the cations. Also, the interaction of the ionic states of the B-site ions with the O$2p$ states is a determining factor of the structure.

The BO6 octahedra consists of six B-O bonds: four B-O$_p$ and two B-O$_a$ bonds. In the *Pbnm* phase the four B-Op bonds have two different values. In general, the larger bonds increase while the shorter bonds decrease with increasing Mn content in the *Pbnm* phase [Fig. 5(b)]. However, there is a disagreement to this trend for $x=0.375$. On the other hand, the two B-Oa bonds are equal and decrease uniformly with increasing Mn-content. After the phase transition, all six B-O bonds merge to a unique bond length in the $R\bar{3}c$ phase. The average <B-O> bond lengths decrease with increasing $x$ [Fig. 5(b)]. Similarly, the average <B-O$_p$> and <B-O$_a$> bond lengths decrease with increasing $x$.

The top face of the pseudocubic structure, La$_1$La$_2$La$_5$La$_6$, is of the shape of a kite [48]. It comprises of two distinct sets of similar adjacent La-La bonds for the *Pbnm* phase and all similar types of La-La bonds for $R\bar{3}c$ phase. The bond set La$_1$-La$_6$ = La$_5$-La$_6$ is larger than the smaller set La$_1$-La$_2$ = La$_5$-La$_2$. Hence, La$_6$ is related to the two larger bonds while La$_2$ is related to the two smaller bonds for the *Pbnm* phase [Fig. S4(c)]. The bond angle ∠La$_1$-La$_6$-La$_5$ < ∠La$_1$-La$_2$-La$_5$, while, ∠La$_2$-La$_1$-La$_6$ = ∠La$_2$-La$_5$-La$_6$. The angles ∠La$_1$-La$_6$-La$_5$ and ∠La$_1$-La$_2$-La$_5$ reduce with increasing Mn-content while ∠La$_2$-La$_1$-La$_6$ and ∠La$_2$-La$_5$-La$_6$ increases. Hence, the shape of the kite becomes sharper with increasing Mn content in the *Pbnm* phase [Fig. S4 (a)]. This indicates a significant La$_8$O$_6$ cage distortion in the *Pbnm* phase. However, for the $R\bar{3}c$ phase ($0.625 \leq x \leq 0.75$), the bond angle ∠La$_1$-La$_6$-La$_5$ = ∠La$_1$-La$_2$-La$_5$, while, ∠La$_2$-La$_1$-La$_6$ = ∠La$_2$-La$_5$-La$_6$, but ∠La$_1$-La$_2$-La$_5$ > ∠La$_2$-La$_5$-La$_6$. Hence, this face is of the form of a parallelogram. This somewhat reveals a lesser distortion of La$_8$O$_6$ cage as compared to *Pbnm* phase. With increasing Mn-content, ∠La$_1$-La$_2$-La$_5$ reduces while ∠La$_2$-La$_5$-La$_6$ increases. This somewhat transforms the quadrilateral to a square-like form resembling a lesser distortion.

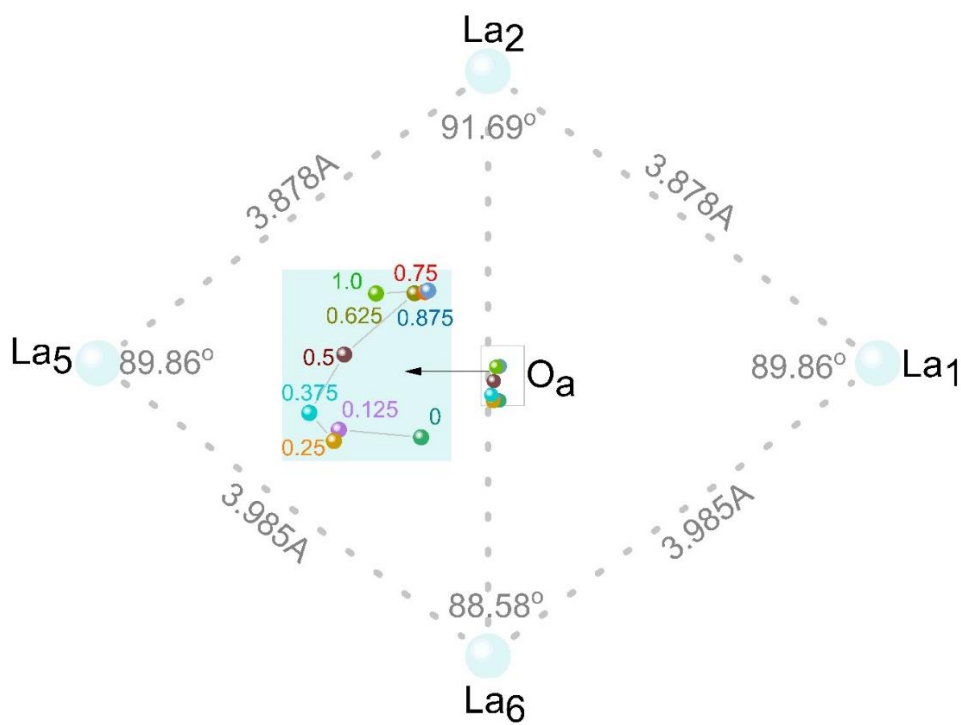

**Fig. 6**: The shifting locus of the $O_a$ in the La-kite structure

An interesting observation can be made from the four La-$O_a$ bond lengths; where, at x~0.25 one observes a changing trend in the values. One of the larger bonds starts to decrease while the other increases with increase of Mn-content until x=0.25 or 0.375. The smaller bonds also seem to follow the pattern almost in the same trend. The average <La-$O_a$> bond length continuously decreases with substitution. One of the larger and smallest bonds become equal after the phase transition and is equal to <La-$O_a$> [Fig. 5(c)]. The other larger bond keeps shortening while the other smaller bond continues to increase until they change trend after x=0.375.

A very similar trend of changing behavior at x=0.25 or 0.375 is observed for the four La-Op bonds [Fig. 5(d)]. The two larger bonds reduce and transform into the smallest with substitution while the smallest ones become the largest. This happens in the *Pbnm* phase. Two bonds of intermediate lengths merge after the phase transition and become equal to the average bond length. However, the smallest becomes the largest and vice versa in the $R\bar{3}c$ phase.

The apical O is not located symmetrically. This is revealed from the distinct La-$O_a$ bond lengths of the top face $La_1La_2La_5La_6$, of both *Pbnm* and $R\bar{3}c$ phase. The difference in $O_a$-$O_p$ bond lengths further supports this claim. This represents a distorted $BO_6$ octahedra. The locus of the $O_a$ ion on the horizontal face is estimated (Fig. 6) [48]. The bond length $La_6$-$O_a$ increases in magnitude with Mn content for $x \leq 0.625$ and thereafter a nominal change is observed for $0.625 \leq x \leq 1.0$ (Fig. S5). The angle of deviation (Fig. S5) decreases for $x \leq 0.375$ and then increases for $0.375 \leq x \leq 0.875$ followed by a nominal decrement for $x=1.0$. It has been proved that the angle of deviation (locus) (Fig. S5) and tilting angle (Fig. 7(c)) of octahedra are correlated and follow the same trend in this present study too [48].

### D. Structural Distortion

The tolerance factor, $t$, is expressed as: $t = (r_{La\_} + r_O)/\sqrt{2}(r_B + r_O)$, where, $r_{La}$, $r_O$, and $r_B$ are the ionic radii of La, O and the B-site ions. Lattice distortion is correlated to '$t$'. Lattice distortion decreases when $t \rightarrow 1$ [Fig.7(c)]. The values reported in literature for both $t(Pbnm)$ [20] and $t(R\bar{3}c)$ [19] are comparable to those obtained in this work for the two phases. The distortion can also be defined by estimating a factor related to the octahedral flatness. A compressed octahedron will have a larger area and a reduced height while an elongated one will see a reverse scenario. Hence, the octahedral flatness can be estimated using the formula: $f = A/h$ where $f$ is the flatness, $A$ is the planar area and $h$ is the height). Flatness decreases drastically for $x \leq 0.125$ with increasing Mn-content. For higher Mn-content $f$ increases again for $0.25 < x \leq 0.50$. At the phase transition there is again a sudden decrease of $f$ followed by a nominal increase with substitution [Fig.7(a)]. On the other hand the octahedral volume decreases continuously irrespective of the phase transition [Fig.7(b)].

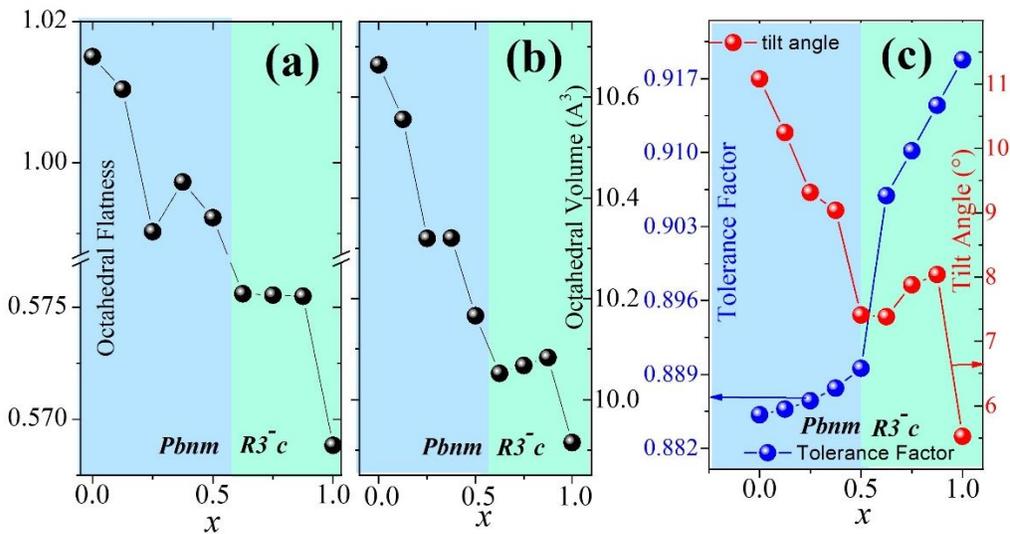

**Fig. 7**: Variation of (a) octahedra flatness (b) octahedra volume with substitution (c) continuous increment of tolerance factor with substitution and continuous decrease in tilt angle is observed for $x \leq 0.5$ and slight increment $0.625 \leq x \leq 0.875$ and with further decrement in tilt angle for $x = 1.0$.

### E. Phonons and Raman Spectra

From the above structural discussion of the materials, it has been established that the chemical substitution influences the structural parameters like bond length/angle, tilting angle, crystal symmetry etc.. These modifications are important enough to bring in changes in the vibrational properties of the constituent ions. To study these changes, Raman spectroscopy has been performed to explore how the phonon modes get modified due to chemical substitution. The Raman spectra (Fig. S6) were fitted using multiple peaks corresponding to known phonon modes. Lattice dynamics of the samples due to Mn substitution was theoretically studied using Density Functional Theory (DFT). Experiment values of the phonon frequencies were compared with simulated results from the DFT calculations.

Group theory analysis reveals twenty four Raman active modes in orthorhombic LFO (space group *Pbnm*) [86]. These can be represented as $\Gamma = (7A_g + 7B_{1g} + 5B_{2g} + 5B_{3g})$. These Raman modes can be further classified into four groups: A-site vibrations (A), oxygen tilting (T), oxygen bending (B), and oxygen stretching (S) [87]. Due to the large mass of La ions, its contribution in phonon modes are present below 155 cm$^{-1}$.

For the LMO-like rhombohedral $R\bar{3}c$ phase [88], the $2a$ positions are occupied by the La atoms which generate four $\Gamma$-point phonons ($1A_{2g}$, $1E_g$, $1A_{2u}$, $1E_u$) [89-92]. The $2b$ positions are occupied by Mn atoms which produce four more phonons ($1A_{1u}$, $1A_{2u}$, $2E_u$) [89-92]. The $6e$ positions are occupied by the O atoms which produce twelve phonons ($1A_{1g}$, $2A_{2g}$, $3E_g$, $1A_{1u}$, $2A_{2u}$, $3E_u$) [89-92]. However, out of the entire possibilities, $\Gamma = A_{1g}+2A_{1u}+3A_{2g}+4A_{2u}+4E_g+6E_u$, the Raman active phonons are limited to ($A_{1g}+3A_{2g}+4E_g$) [91].

Experimentally, phonons are observed to shift and broaden with increasing Mn content. The phonon modes are rarely visible for LMO. Phonon energies modify with composition and depend on the phases. For $x>0.75$, the peak intensity weakens considerably. Although the phonon modes are different in the two phases a continuity of the shifting needs attention on nature of the vibrations of these two phases. Hence, individual vibrations were studied using the DFT displacement pattern. Eigenvectors of the phonon modes obtained from DFT calculations gave a better insight into the detailed lattice dynamics [48, 93, 94]. The calculated and experimental Raman frequencies were compared. The values matched well (Table 1 (a and b)). A correlation with the structural parameters like bond-lengths and bond angles and their influence on the Raman modes is proposed. The high frequency broad Raman bands in the range > 400cm-1 correspond to the dynamic Jahn-Teller distortion ($Mn^{3+}$ sites) [95].

**Table 1**: The calculated and experimental Raman frequencies of LFMO sample: (a) *Pbnm* phase ($0 \le x \le 0.5$) (b) $R\bar{3}c$ phase ($0.625 \le x \le 1.0$)

| 1(a) | Exp_0 | Cal_0 | Exp_0.125 | Exp_0.25 | Cal_0.25 | Exp_0.375 | Exp_0.5 | Cal_0.50 |
|---|---|---|---|---|---|---|---|---|
| Ag(1) | 79.91 | 68.7 | 74.14 | 74.1 | 63.4 | 74.20 | 76.81 | 62.6 |
| B1g(1) | 97.98 | 96.4 | 85.63 | 85.43 | 93.2 | 85.2 | 88.30 | 90.5 |
| B2g(1) | 102.8 | 109.7 | 102.94 | 102.90 | 107.8 | 106.00 | 106.4 | 101.1 |
| B1g(2) | 134.2 | 125.8 | 129.45 | 129.4 | 115.4 | 143.55 | 139.4 | 111.6 |
| Ag(2) | 139.9 | 133.0 | 140.26 | 140.9 | 139.4 | 151.2 | 170.1 | 133.7 |
| B3g |  | 147.5 |  |  | 140.8 |  |  | 144.4 |
| B1g(3) | 158.41 | 153.6 | 166.34 | 169.0 | 153.1 | 170.0 | 200.0 | 146.1 |
| B2g(2) | 181.2 | 164.4 | 175.55 | 175.53 | 158.1 | 173.78 | 209.18 | 164.4 |
| Ag(3) | 183.6 | 171.7 | 255.2 | 261.8 | 169.7 | 185.67 | 227.5 | 168.9 |
| Ag(4) | 267.85 | 225.0 | 279.2 | 279.28 | 226.4 | 192 | 290.27 | 229.9 |
| Ag |  | 287.0 |  |  | 273.3 |  |  | 265.3 |
| B3g(1) | 295.3 | 288.8 | 292.2 | 289.9 | 285.4 | 212.7 | 344.97 | 284.9 |
| B2g(3) | 310.27 | 345.6 | 325.0 | 339.0 | 342.5 | 262.22 | 359.0 | 329.9 |
| B1g(4) | 335.1 | 347.6 | 351.9 | 351.99 | 348.0 | 292.50 | 368.71 | 347.0 |
| B1g | 411.69 | 377.8 | 383.9 | 383.9 | 372.2 | 331.8 | 389.8 | 375.2 |
| B3g(2) | 414.0 | 403.1 | 427.71 | 427.6 | 405.3 | 424.04 | 405.77 | 395.9 |
| Ag(5) | 432.5 | 412.0 | 480.72 | 480.6 | 418.1 | 466.9 | 421.1 | 413.1 |
| B1g(5) | 443.2 | 468.9 | 519.3 | 519.2 | 473.4 | 487.18 | 487.2 | 471.3 |
| B3g(3) | 493.69 | 561.0 | 544.11 | 544.11 | 564.9 | 548.18 | 533.85 | 571.0 |
| B2g |  | 562.9 |  |  | 566.4 |  |  | 574.6 |
| Ag(6) | 568.41 | 565.9 | 582.4 | 582.48 | 578.3 | 550.2 | 567.4 | 580.1 |
| B2g(4) | 589.24 | 595.2 | 594.53 | 594.53 | 593.7 | 604.9 | 592.8 | 602.0 |
| B1g(6) | 627.08 | 620.1 | 618.87 | 618.6 | 624.5 | 636.27 | 632.00 | 624.2 |
| B3g(4) | 645.92 | 650.7 | 642.95 | 642.83 | 647.5 | 657.72 | 667.74 | 642.5 |

| 1(b) | Exp_0.625 | Exp_0.75 | Cal_0.75 | Exp_0.875 | Exp_1.0 | Cal_1.0 |
|---|---|---|---|---|---|---|
| Eg(1) | 90.5 | 90.64 | 82.2 | 90.8 | 88.81 | 13.1 |
| A2g(1) | 113.2 | 115.0 | 113.0 | 115.2 | 115.6 | 128.9 |
| Eg(2) | 156.9 | 156.0 | 155.9 | 155.0 | 165.0 | 155.1 |
| A1g(1) | 212.6 | 218.7 | 233.8 | 235.4 | 231.13 | 244.2 |
| A2g(2) | 324.06 | 319.4 | 339.8 | 315.54 | 320.01 | 368.3 |
| Eg(3) | 507.06 | 504.7 | 565.9 | 502.5 | 518.48 | 395.4 |
| Eg(4) | 640.03 | 638.9 | 617 | 637.11 | 640.0 | 600.6 |
| A2g(3) | 669.8 | 668.09 | 662.8 | 666.0 | 668.8 | 669.3 |

The *Pbnm* → $R\bar{3}c$ structural phase transition is most probably on account of the change in B-site ionic radius due to Fe ions being replaced by Mn ion which promotes a rotation or distortion of the BO6 octahedra. There are eight $R\bar{3}c$ modes which are similar in vibrational symmetries with some *Pbnm* modes and these have comparable energies. These modes can be calculated from DFT simulations and can be enlisted below (*Pbnm* → $R\bar{3}c$).

*B1g*(1) (96.4 cm$^{-1}$) → *Eg(1)* (13.1 cm$^{-1}$); *B2g*(1) (109.7 cm$^{-1}$) → *A2g*(1) (128.9 cm$^{-1}$); *B1g*(3) (153.6 cm$^{-1}$) → *Eg(2)* (155.1 cm$^{-1}$); *B2g*(2) (164.4 cm$^{-1}$) → *A1g*(1) (244.2 cm$^{-1}$); *B2g*(3) (345.6 cm$^{-1}$) → *A2g*(2) (368.3 cm$^{-1}$); *B3g*(3) (561.0 cm$^{-1}$) → *Eg(3)* (395.4 cm$^{-1}$); *B1g*(6) (620.1 cm$^{-1}$) → *Eg(4)* (600.6 cm$^{-1}$); and *B3g*(4) (650.7 cm$^{-1}$) → *A2g*(3) (669.3 cm$^{-1}$).

The relative vibrations between the A/B/O ions can be resolved into two perpendicular directions along or perpendicular to a specific direction. For the *Pbnm* phase this specific direction is along the c-axis, i.e. the pseudocubic [001] direction while for the $R\bar{3}c$ phase it is along the rhombohedral [111]$_R$ direction, i.e. the pseudocubic [101] direction.

The *Pbnm* $B_{1g}$(1) mode [Fig. 8.i (a)] at 96.4 cm$^{-1}$ (theoretically calculated for LFO) is tilting vibration of the BO6 octahedra in a direction nearly parallel to the vertical planes of a perovskite structure. It is associated with a strong La-ion vibration along a direction perpendicular to [001] direction. Among the four planar O ions attached with B-ion, two move along and the other two move opposite to the direction [001]. The two apical O-ions move perpendicular to the [001] direction. This results in the tilting vibrations of the BO6 octahedra along the [110] direction. On the other hand, in a general $R\bar{3}c$ phase, the tilting of the BO6 octahedra ~55° w.r.t. the [111] axis. The theoretically calculated $R\bar{3}c$ *Eg(1)* mode for LMO at 13.1 cm$^{-1}$ has a similar vibration pattern as the *Pbnm* *B1g*(1) mode with a very similar octahedral tilting and a La displacement. However, the La-ion vibration seems to be weaker in LMO than LFO. From the structural analysis it was observed that the La-Oa bonds are strongly affected by the substitution. Note that the experimental phonon frequency of these *Pbnm* *B1g*(1) and the $R\bar{3}c$ *Eg(1)* modes bears an apparent correlation with the La-O bond lengths. From experiment, the *Pbnm* *B1g*(1)phonon frequency redshifts from 97.98cm$^{-1}$ (theoretical 96.4 cm$^{-1}$) in LFO, 85.63 cm$^{-1}$ in LFMO12, 85.43 cm$^{-1}$ (93.2 cm$^{-1}$) in LFMO25, to 85.2 cm$^{-1}$ in LFMO37. However, for LFMO50 one observes a slight blue shift to 88.30 cm$^{-1}$ although the theoretically calculated value still shows a red shift to 90.5 cm$^{-1}$. This feature is a hint towards an approaching phase transition in the material. Note that the structural phase transition happened for LFMO62. The phonon frequencies observed reveal a mild blue shift from 90.5 cm$^{-1}$ in LFMO62, 90.64 cm$^{-1}$ (theoretical 82.2 cm$^{-1}$) in LFMO75, to 90.8 cm$^{-1}$ in LFMO87. However, for LMO there is a nominal red shift to 88.81 cm$^{-1}$ in agreement with a much drastic theoretically calculated value of 13.1 cm$^{-1}$.

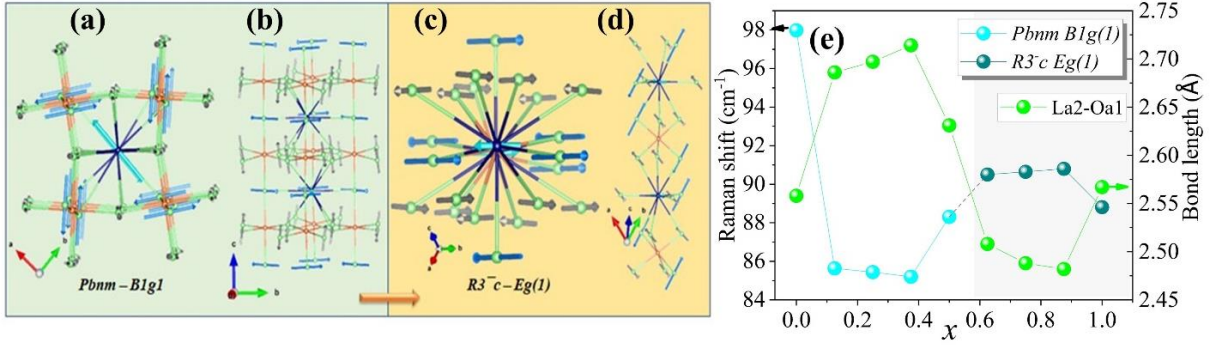

**Fig. 8.i**: (a) Modification of *Pbnm* phonon mode ((a) top view and (b) side view) $B1g(1) \rightarrow Eg(1)$ $R\bar{3}c$ phonon mode ((c) top view and (d) side view) (b) Strong correlation between Raman shift and La$_2$-O$_{a1}$ bond length.

Similar to these modes a similarity between the trends between La-O bond lengths and phonon frequencies are observed between the correlated couples [tilting modes *Pbnm* $B2g(2)$ and $R\bar{3}c$ $A1g(1)$ [Fig. S7.i (a)], [antisymmetric stretching modes *Pbnm* $B3g(3)$ and $R\bar{3}c$ $Eg(3)$ [Fig. S7.ii (a)] and [breathing modes *Pbnm* $B3g(4)$ and $R\bar{3}c$ $A2g(3)$ (Fig. S7.iii)].

The *Pbnm* $B_{2g}(1)$ mode at 109.7 cm$^{-1}$ corresponds to a rotational vibration of the BO6 octahedron along with extremely weak La vibrations. The adjacent La-ions belonging to neighboring octahedra vibrate in opposite directions. The experimental phonon frequency recorded was for LFO ~102.8 cm$^{-1}$ (theoretical 109.7 cm$^{-1}$), for LFMO12 ~102.94 cm$^{-1}$, LFMO25 ~102.90 cm$^{-1}$ (theoretical 107.8 cm$^{-1}$), LFMO37~106.00 cm$^{-1}$, LFMO50 106.4 cm$^{-1}$ (theoretical 101.1 cm$^{-1}$). Hence, not much change is observed in the phonon frequency of this mode, both experimentally and theoretically. However, such a vibration concerning octahedral rotation is not observed. The closest phonon mode to such frequencies is the $R\bar{3}c$ $A_{2g}(1)$ mode (128.9 cm$^{-1}$). This mode has extremely negligible O ion movement, thereby nullifying the octahedral variance. However, the vibration of the La-ion along [111]$_R$ direction becomes much stronger with adjacent La ions vibrating in opposite directions. An interesting correlation is observed between the La-B ion-separations and the phonon frequency of these two modes [Fig. 8.ii]. It seems from this correlation that the La-B separation intervenes the rotational motion of the octahedra, thereby letting the La-ion vibration be the principal determining factor of the phonon energy. Some La-O bond strengths are also similar to such trends.

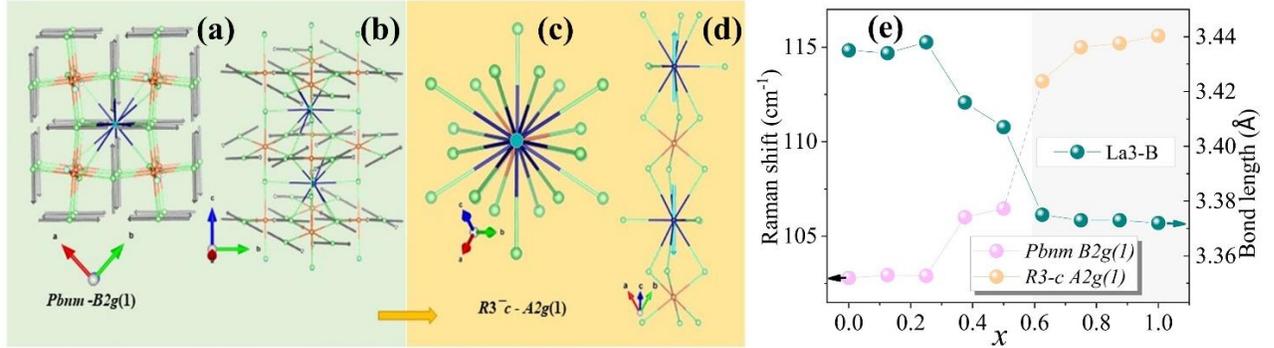

**Fig. 8.ii**: (a) Modification of *Pbnm* phonon mode ((a) top view and (b) side view) $B2g(1) \rightarrow A2g(1)\ R\bar{3}c$ phonon mode phonon mode ((c) top view and (d) side view). ii(b) Strong correlation between Raman shift and La$_3$-B bond length.

Similar to the above discussion correlation between phonon frequencies and different bond lengths (Fig. S7.i to Fig. S7.vi) were observed in this study. It seems that with the incorporation of Mn in place of Fe, the structure gets modified restricting certain atom vibrations while promoting some, thereby requiring a *Pbnm* → $R\bar{3}c$ transition.

**F. Electronic Band gap properties:**

The room temperature UV–Vis absorption spectra [Fig. 9(a)] of all the samples was used to plot the Tauc plots and thereby estimate the optical band gap, $E_g$, from Tauc relationship *(αhv) = A (hv − $E_g$)$^n$,* where $\alpha$ = absorption coefficient, $hv$ = photon energy, *A* is a constant, and *E* is the bandgap [96]. The exponent $n = ½$ for direct and 2 for indirect bandgap [97, 98]. Bandgap plot (Fig. 9(b)) reveals that bandgap continuously decreases with Mn substitution from 2.49eV in LFO to 1.35eV in LMO. The values of bandgap for LFO [99] and LMO [6] are comparable to literature. Total DOS for LFO and LMO i.e. two extreme end members of LFMO have been calculated to see how the electron density of states are distributed near the Fermi level which participates in conduction process. It can be observed from the DOS plot (Fig S9) that the gap near the Fermi level is more in LFO than LMO. Bandgap of LFO was measured to be ~1.2 and for LMO is ~0.994.

Lattice strain was found to increase from XRD data with Mn substitution. Such a strain may create electronic defect states between the band gap. Existence of such defect states related to disorder in the system can be estimated by calculating the Urbach energy [100]. Urbach energy was found to increase with Mn substitution (Fig. 9(b)).

It is evident from Fig. S4 (b) that the average ∠B-O-B increases with increasing Mn-content from 157.835° to 168.95°. This may be the result of a more relaxed structure with lesser distortions as has been observed from structural studies. This lessening of distortion can be associated with a decrease in the bandgap with increasing Mn-content [101].

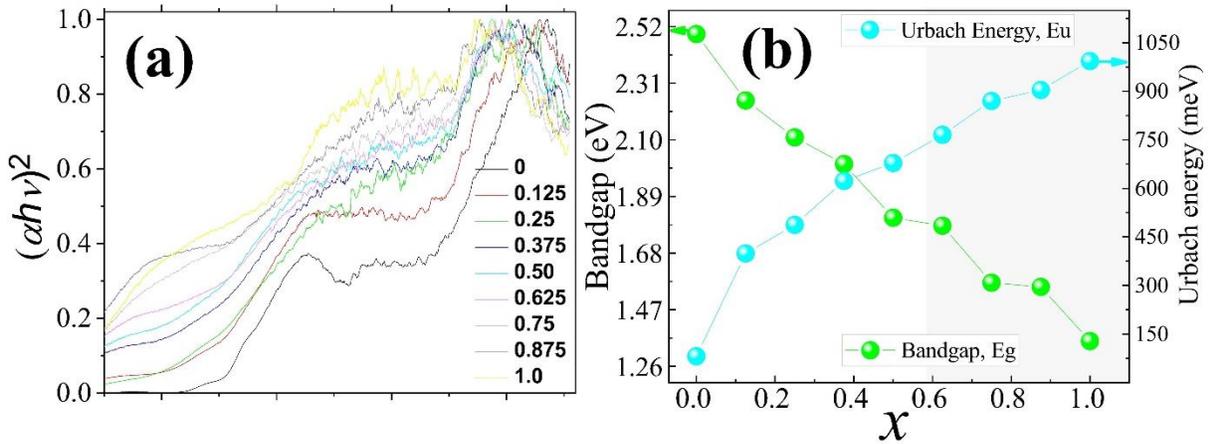

**Fig. 9:** (a) The Tauc plots of the UV-Vis spectra of LFMO sample (b) decrement of electronic band gap and increase of Urbach energy with Mn content.

The theoretical magnetic moments for the $Fe^{3+}$ and the $Mn^{3+}$ ions are 5 Bohr magneton and 4 Bohr magneton respectively. The different d- electron environments in $FeO_6$ and $MnO_6$ octahedral units leads different crystal field splitting between $t_{2g}$ and $e_g$ levels which affects the band gap between two electronic states ie oxygen $2p$ valence band and Fe/Mn $t_{2g}$ conduction band. From electronic properties studies, one can observe two strong absorption edges. A valence band (VB) → B $t_{2g}$ minority spin states excitation gives rise to the first absorption edge, while a VB → B $e_g$ minority spin states is responsible for the second [99].

**Conclusions:**

Rietveld analysis of room temperature XRD data of the LFMO confirms a phase transition from a pure orthorhombic (*Pbnm*) phase (0≤ $x$ ≤ 0.50) to a rhombohedral (*R3¯c*) phase (for $0.625 \leq x \leq 1.0$). All the lattice parameters a, b and c decrease with increasing Mn content, with the exception that c increases in the $R\bar{3}c$ phase. The tilting of the BO6 octahedra determines the position of the apical oxygen. The position modifies as a result of an extra amount of oxygen incorporated due to probable $Mn^{4+}$ substitution. Different phonon modes can be correlated with different bond lengths or angles. Zone center phonon calculations reveal sensitive correlations between phonon frequencies, local distortion and bond length. Blue shift of phonons was observed for most of the prominent modes. The rotational $A_{2g}$ phonon may be correlated to the $Pbnm \rightarrow R\bar{3}c$ phase transition. XPS results reveal a mixed presence of $Mn^{3+}$ in both high and low spin states along with a considerable amount of $Mn^{4+}$ ions. On the other hand the Fe is in a mixed high/low spin $Fe^{3+}$ state. The extra charge of $Mn^{4+}$ may also be a factor in the structural properties and hence modify the functional implications. The optical band gap decreases with increasing Mn content most probably due to different d- electron environments in $FeO_6$ and $MnO_6$ octahedral units leading to variations in the crystal field splitting between $t_{2g}$ and $e_g$ levels. This is manifested in the $BO_6$ octahedral flattening. The structural arrangement controls the octahedral distortion/tilting and plays a key role in tuning the phase and electronic properties of LFMO.

**Acknowledgements**

The first author and corresponding author acknowledge the Department of Science and Technology (DST, Govt. of India) FIST program for providing a Horiba Raman spectrometer to the Discipline of Physics at IIT Indore, employed to gather high quality Raman spectra. They also thank the Ministry of Education and Indian Institute of Technology Indore for providing research infrastructure. The authors are highly indebted to Prof. M. K. Glazer, Oxford University for his immense support with Glazer representation of the octahedral tilting in the samples. The first author acknowledges the DST for financial support under the Women Scientist Scheme-A (SR/WOS-A/PM-99/2016 (G)). All the authors are thankful to the Chairman, Physics Department KFUPM, KSA for the XPS facility.

# *Pbnm* to $R\bar{3}c$ phase transformation in *(1-x)*LaFeO$_3$*x*LaMnO$_3$ solid solution due to modifications in structure, octahedral tilt and valence states of Fe/Mn

E. G. Rini[1], Mayanak. K. Gupta[2], R. Mittal[2,3], A. Mekki[4], Mohammed H. Al Saeed[4], Somaditya Sen[1*]

[1]*Department of Physics, Indian Institute of Technology Indore, Indore, 453552, India*

[2]*Solid State Physics Division Bhabha Atomic Research Centre Mumbai 400085, India*

[3]*Homi Bhabha National Institute, Anushaktinagar, Mumbai, 400094, India*

[4]*Department of Physics, King Fahd University of Petroleum & Minerals Dhahran, 31261, Saudi Arabia*

*\* Corresponding author: sens@iiti.ac.in (SS)*

**Supporting information:**

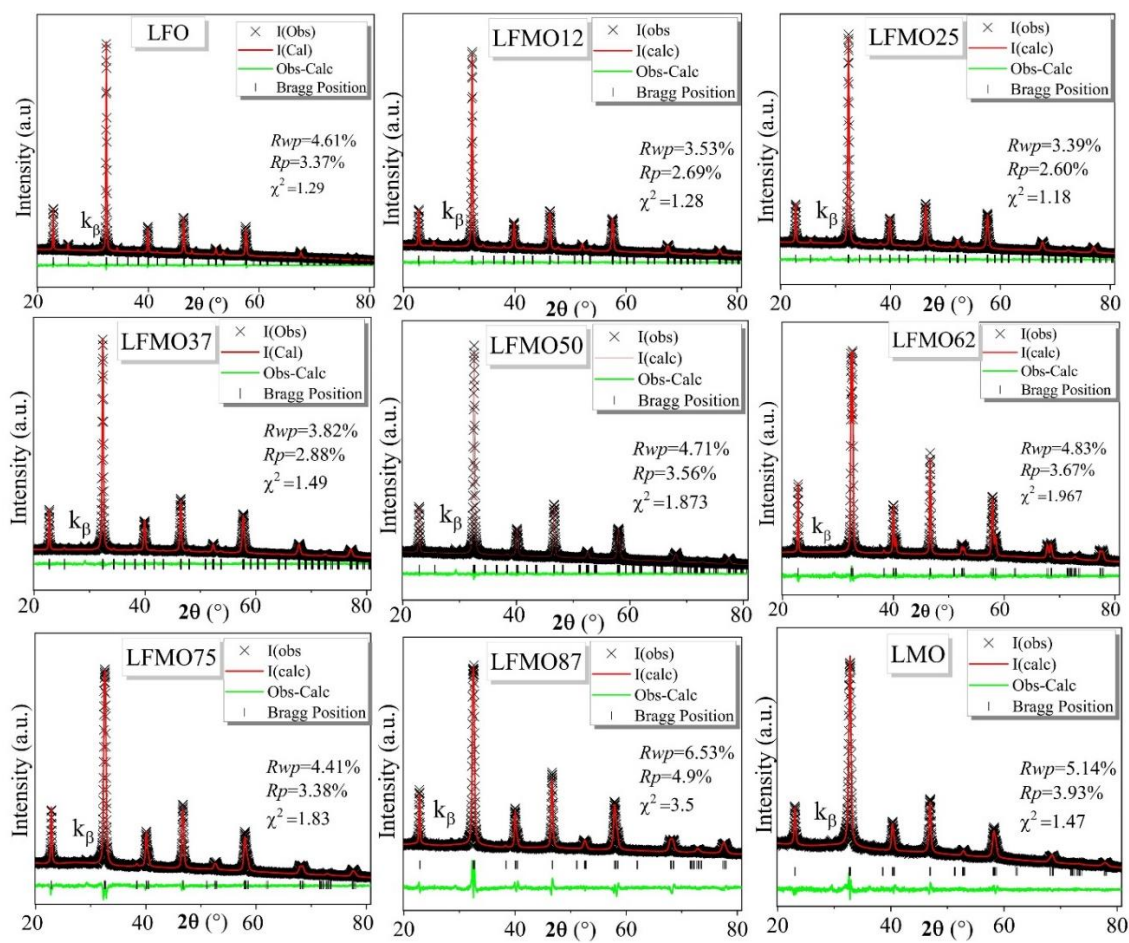

**Fig. S1:** Rietveld analysis of XRD patterns showing excellent fit between theory and experiments revealing a *Pbnm* (0 ≤ $x$ ≤ 0.5) and $R\bar{3}c$ (0.625 ≤ $x$ ≤ 1.0) for all samples.

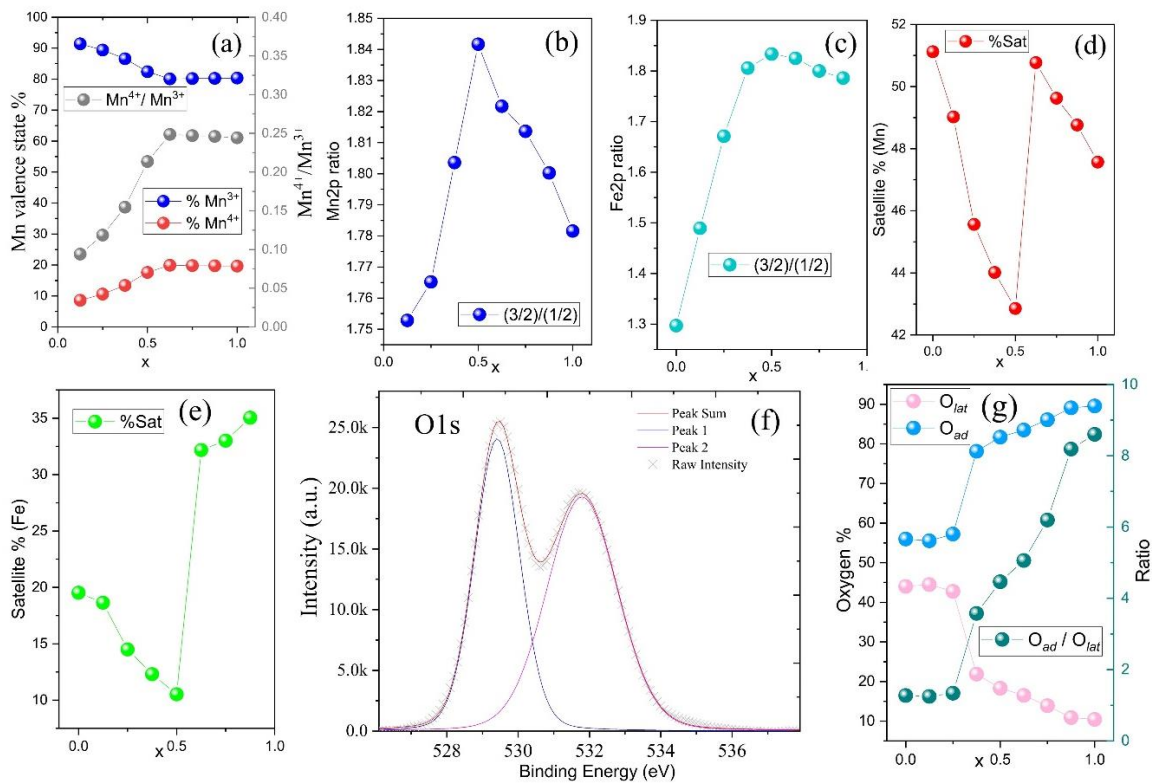

**Fig. S2.i:** (a) Variation of % Mn valance state with composition, $x$; (b) Mn2p ratio; (c) Fe2p ratio; (d) Mn Satellite %; (e) Fe Satellite %; (f) The deconvolution of O1s (peak 1 and peak 2 corresponds to $O_{lat}$ and $O_{ad}$ respectively) and (g) Oxygen %.

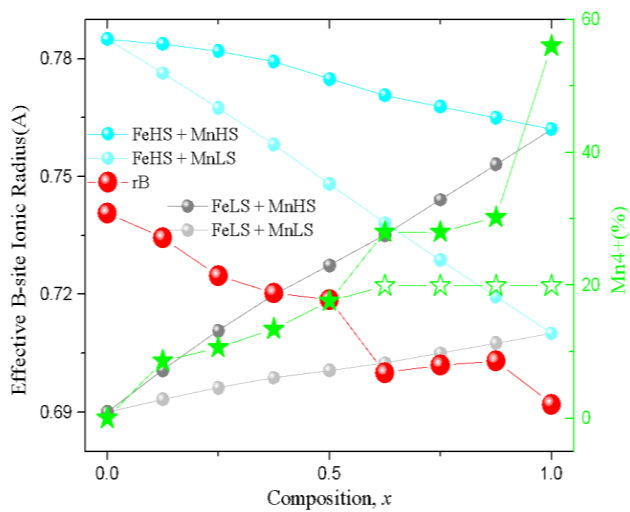

**Fig. S2.ii:** The theoretical content of $Mn^{4+}$ (solid green star) with different combinations of $Fe^{3+}$ and $Mn^{3+}$ spin states that can replicate effective B-site ionic radius obtained from XRD analysis and hollow green star is observed $Mn^{4+}$ % obtained from XPS analysis.

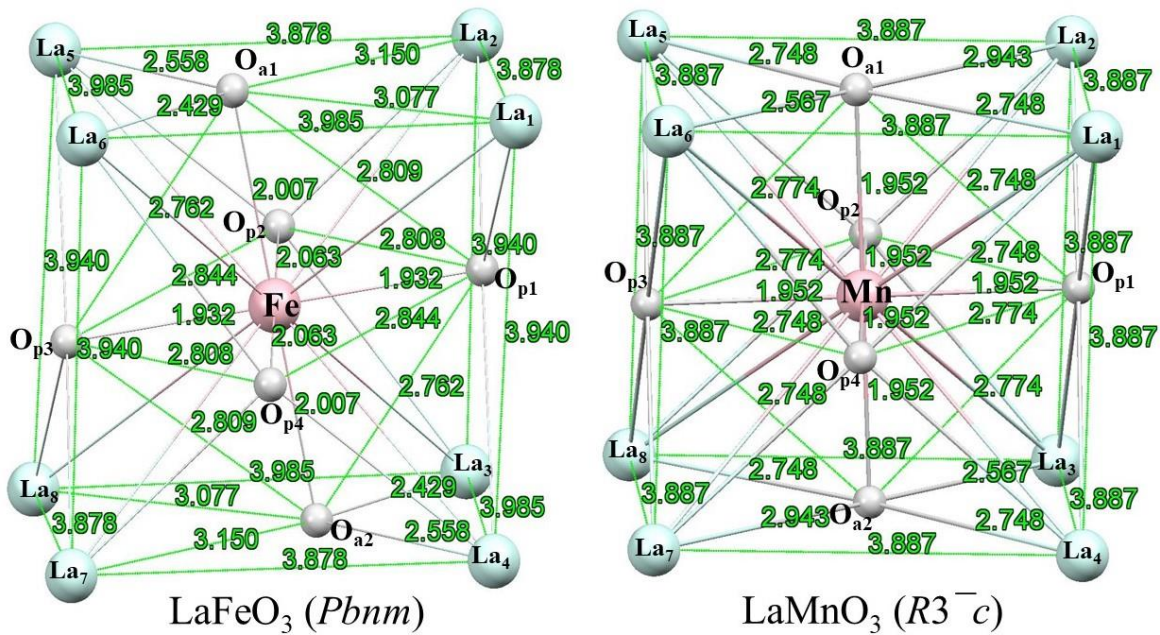

**Fig. S3:** A pseudocubic perovskite unit containing two components: a $BO_6$ octahedra and a $La_8O_6$ cuboidal cage [48] as obtained from Rietveld refined CIF files using Mercury software

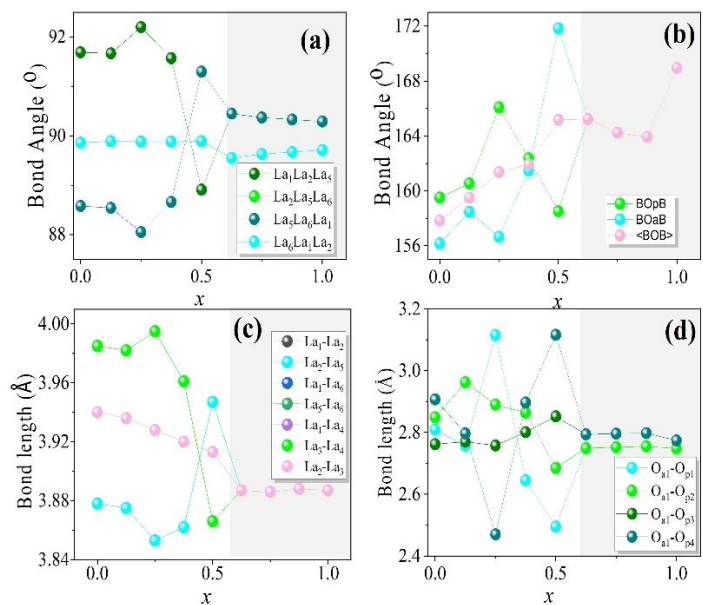

**Fig. S4:** Variation of bond length and bond angles with composition, *x*.

(a) ∠La-La-La bond angle (b) ∠B-O-B bond angle (c) La-La bond length (d) O-O bond length

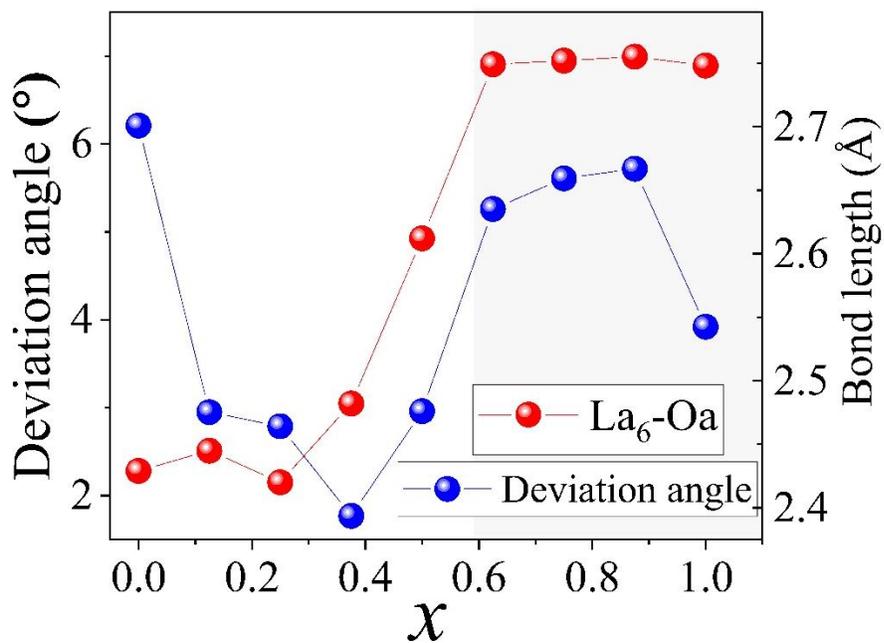

**Fig. S5:** The variation of deviation angle (∠OaLa$_6$La$_2$) and La$_6$-Oa bond length with composition, *x*.

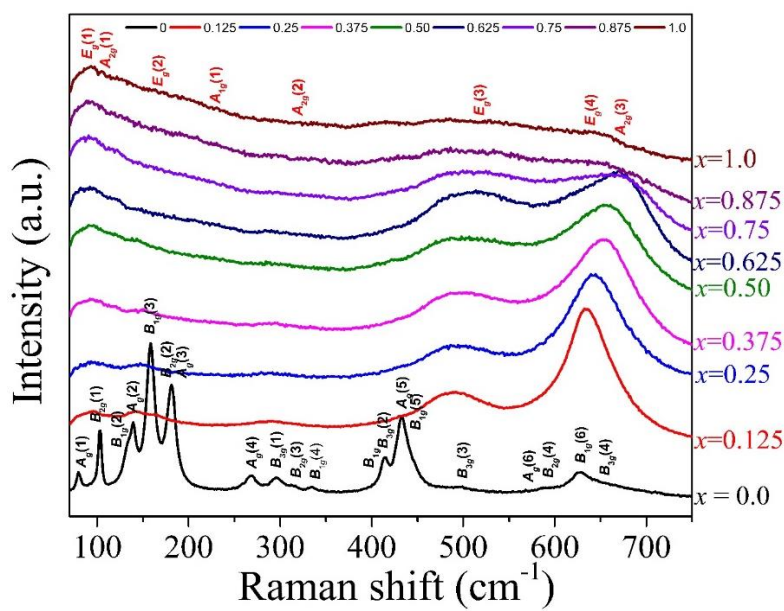

**Fig. S6:** Shifting and broadening of Raman modes observed with increasing substitution for LFMO nanoparticles. The peak intensity weakens for x>0.75.

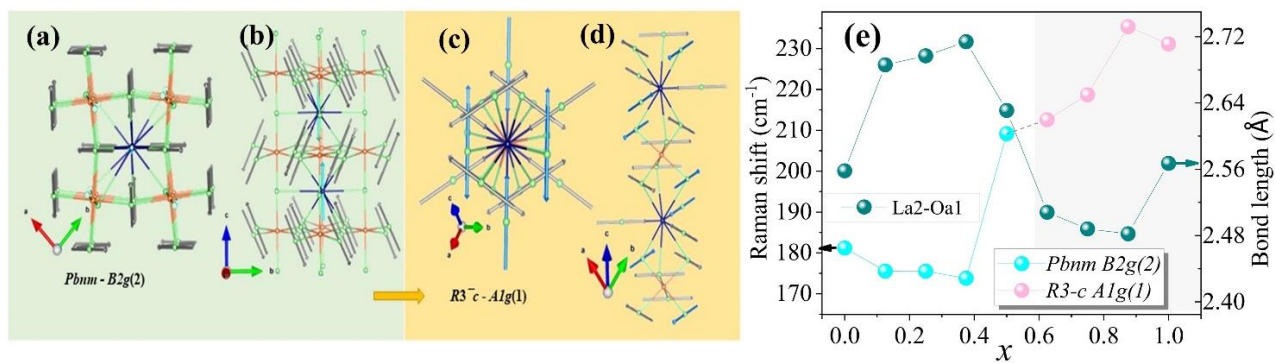

**Fig. S7.i:** (a) Modification of *Pbnm* phonon mode ((a) top view and (b) side view) $B2g(2) \rightarrow A1g(1)$ $R\bar{3}c$ phonon mode ((c) top view and (d) side view). ii(b) Strong correlation between Raman shift and $La_2$-$O_{a1}$ bond length.

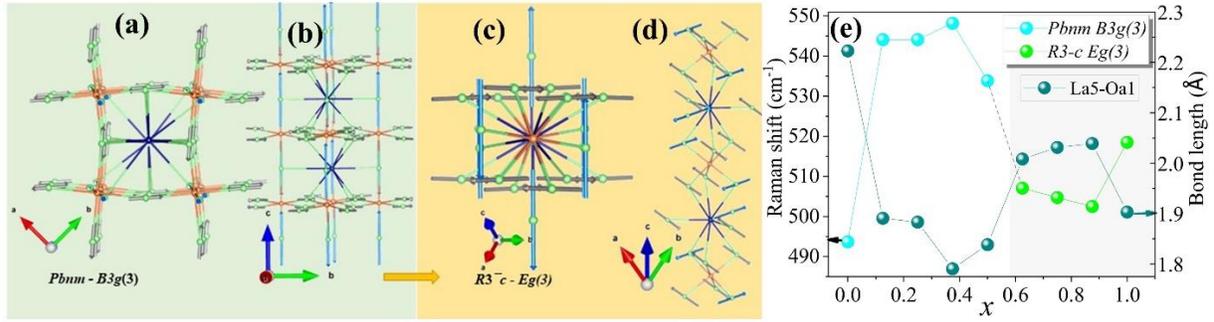

**Fig. S7.ii:** (a) Modification of *Pbnm* phonon mode ((a) top view and (b) side view) $B3g(3) \rightarrow Eg(3)$ $R\bar{3}c$ phonon mode phonon mode ((c) top view and (d) side view). ii(b) Strong correlation between Raman shift and $La_5$-$O_{a1}$ bond length.

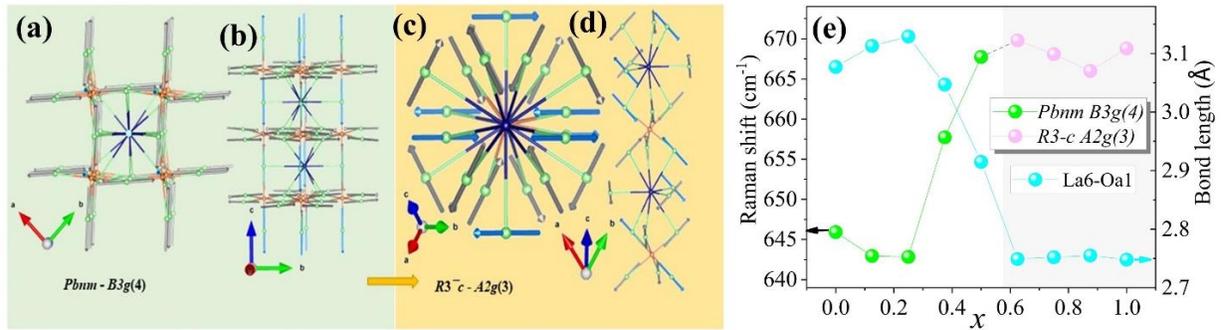

**Fig. S7.iii:** (a) Modification of *Pbnm* phonon mode ((a) top view and (b) side view) $B3g(4) \rightarrow A2g(3)$ $R\bar{3}c$ phonon mode phonon mode ((c) top view and (d) side view)). ii(b) Strong correlation between Raman shift and $La_6$-$O_{a1}$ bond length.

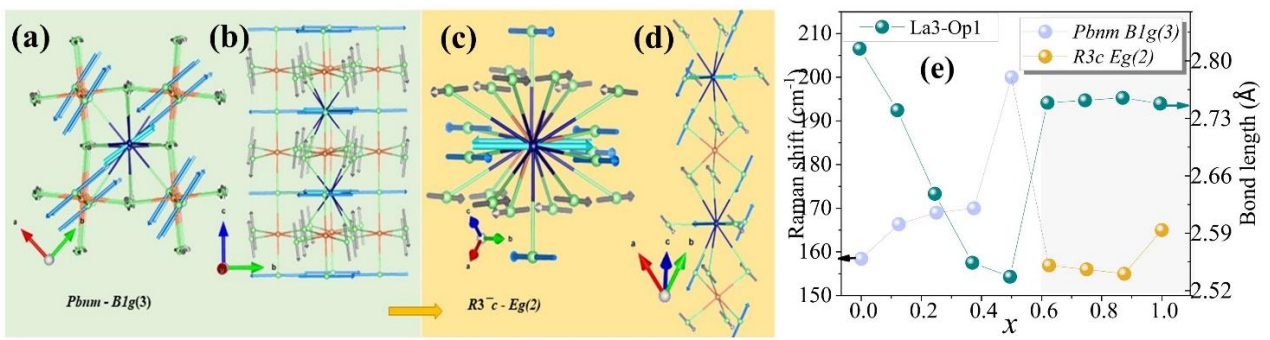

**Fig. S7.iv:** Modification of *Pbnm* phonon mode ((a) top view and (b) side view) $B1g(3) \rightarrow Eg(2)$ $R\bar{3}c$ phonon mode phonon mode ((c) top view and (d) side view). (b) Strong correlation between Raman shift and $La_3$-$O_{p1}$ bond length.

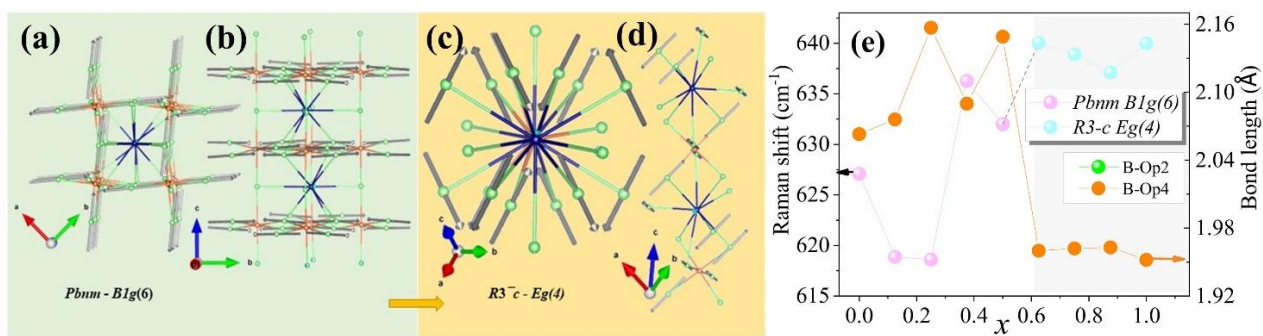

**Fig. S7.v:** Modification of *Pbnm* phonon mode ((a) top view and (b) side view) *B1g*(6)→*Eg(4)* $R\bar{3}c$ phonon mode phonon mode ((c) top view and (d) side view). (b) Strong correlation between Raman shift and B-$O_{p2}$ and B-$O_{p4}$ bond lengths.

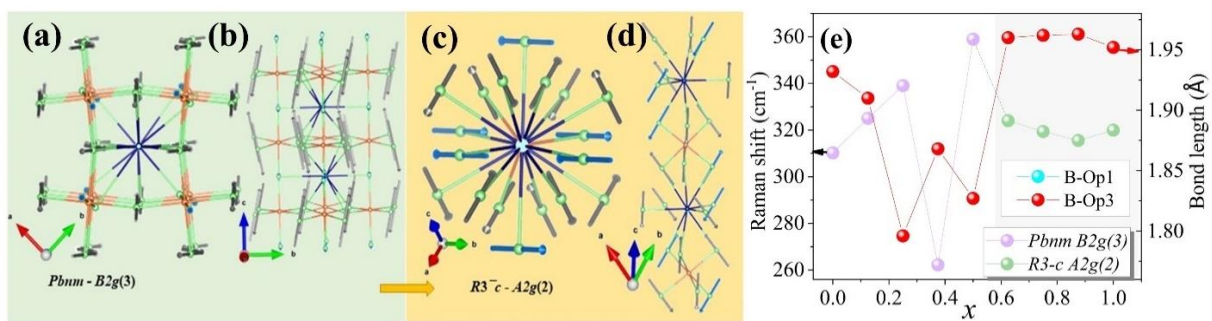

**Fig. S7.vi:** (a) Modification of *Pbnm* phonon mode ((a) top view and (b) side view) *B2g*(3)→*A2g*(2) $R\bar{3}c$ phonon mode phonon mode ((c) top view and (d) side view). (b) Strong correlation between Raman shift and B-$O_{p1}$ and B-$O_{p3}$ bond lengths.

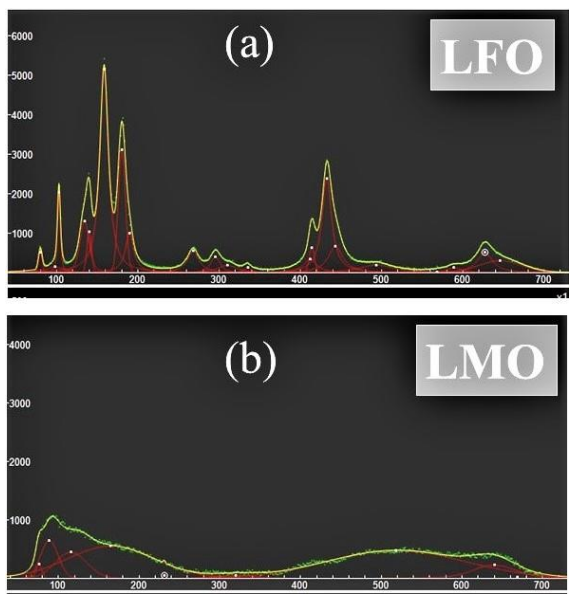

**Fig. S8:** Convolution of Raman spectra using FitYK software. (a) LFO (b) LMO

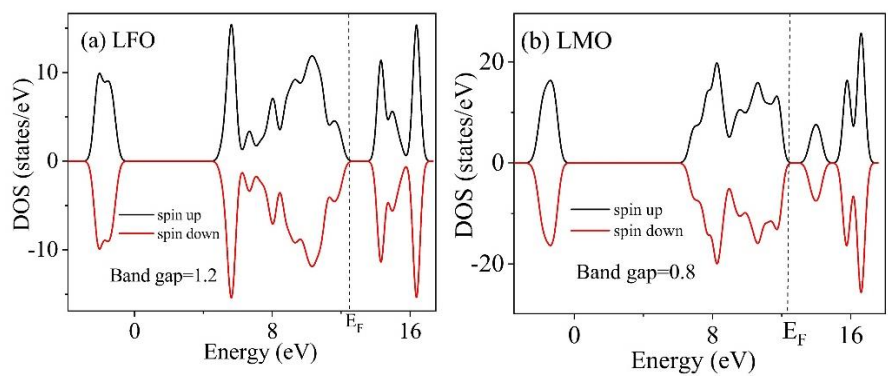

**Fig. S9:** The calculated density of states (DOS) using DFT+U(U=2eV) for (a) LFO and (b) LMO.

**Table S1:** Results from Rietveld refinement analysis for LFMO (a) *Pbnm* ($0 \leq x \leq 0.5$) and (b) $R\bar{3}c$ ($0.625 \leq x \leq 1.0$) samples sintered at 750 °C.

| $x$ (a) | Parameters | | | | |
|---|---|---|---|---|---|
| 0 | a (Å) = 5.55390 (0.00025) b (Å) = 5.56510 (0.00019) c (Å) = 7.85350 (0.00032) | | | | |
| | | $x$ | $y$ | $z$ | $U_{iso}$(Å$^2$) | Occupancy |
| | La | 0.993221 (0) | 0.027712 (0) | 0.250000 (0) | 0.00268 | 1.0 |
| | Fe | 0.000000 (0) | 0.500000 (0) | 0.000000 (0) | 0.00000 | 1.0 |
| | O1 | 0.715939 (0) | 0.267033 (0) | 0.037262 (0) | 0.00208 | 1.0 |
| | O2 | 0.072043 (0) | 0.480506 (0) | 0.250000 (0) | 0.08968 | 1.0 |

| 0.125 | a (Å) = 5.55171 (0.00039) b (Å) = 5.55990 (0.00049) c (Å) = 7.84788 (0.00040) | | | | | |
|---|---|---|---|---|---|---|
| | | $x$ | $y$ | $z$ | $U_{iso}$(Å$^2$) | Occupancy |
| | La | 0.993185 (0) | 0.027230 (0) | 0.250000 (0) | 0.00374 | 1.0 |
| | Fe | 0.000000 (0) | 0.500000 (0) | 0.000000 (0) | 0.00123 | 0.875 |
| | Mn | 0.000000 (0) | 0.500000 (0) | 0.000000 (0) | 0.01334 | 0.125 |
| | O1 | 0.705319 (0) | 0.273099 (0) | 0.026212 (0) | 0.02371 | 1.0 |
| | O2 | 0.067027 (0) | 0.504598 (0) | 0.250000 (0) | 0.02017 | 1.0 |
| 0.25 | a (Å) = 5.54440 (0.00053) b (Å) = 5.55260 (0.00063) c (Å) = 7.83790 (0.00036) | | | | | |
| | | $x$ | $y$ | $z$ | $U_{iso}$(Å$^2$) | Occupancy |
| | La | 0.991705 (0) | 0.021587 (0) | 0.250000 (0) | 0.00571 | 1.0 |
| | Fe | 0.000000 (0) | 0.500000 (0) | 0.000000 (0) | 0.00169 | 0.75 |
| | Mn | 0.000000 (0) | 0.500000 (0) | 0.000000 (0) | 0.00039 | 0.25 |
| | O1 | 0.744581 (0) | 0.300674 (0) | 0.001522 (0) | 0.03661 | 1.0 |
| | O2 | 0.074964 (0) | 0.502380 (0) | 0.250000 (0) | 0.00634 | 1.0 |
| 0.375 | a (Å) = 5.52673 (0.00036) b (Å) = 5.53608 (0.00053) c (Å) = 7.82824 (0.00028) | | | | | |
| | La | 0.993658 (0) | 0.018970 (0) | 0.250000 (0) | 0.01514 | 1.0 |
| | Fe | 0.000000(0) | 0.500000 (0) | 0.000000 (0) | 0.00508 | 0.625 |
| | Mn | 0.000000 (0) | 0.500000 (0) | 0.000000 (0) | 0.00861 | 0.375 |
| | O1 | 0.725295 (0) | 0.245997 (0) | 0.037199 (0) | 0.02440 | 1.0 |
| | O2 | 0.057553 (0) | 0.505095 (0) | 0.250000 (0) | 0.00478 | 1.0 |
| 0.50 | a (Å) = 5.52000 (0.00032) b (Å) = 5.52900 (0.00045) c (Å) = 7.82123 (0.00067) | | | | | |
| | La | 0.005225 (0) | 0.012466 (0) | 0.250000 (0) | 0.01024 | 1.0 |
| | Fe | 0.000000 (0) | 0.500000 (0) | 0.000000 (0) | 0.00019 | 0.50 |
| | Mn | 0.000000 (0) | 0.500000 (0) | 0.000000 (0) | 0.00037 | 0.50 |
| | O1 | 0.724 (6) | 0.234 (7) | 0.0468 (13) | 0.00022 | 1.0 |
| | O2 | 0.022249 (0) | 0.48806 (0) | 0.250000 (0) | 0.02688 | 1.0 |

| $x$ (b) | Parameters |
|---|---|
| 0.625 | a (Å) = 5.51900 (0.00007) b (Å) = 5.51900 (0.00007) c (Å) = 13.35980 (0.00030) |

|  | La | 0.000000 (0) | 0.000000 (0) | 0.250000 (0) | 0.03401 | 1.0 |
|---|---|---|---|---|---|---|
|  | Fe | 0.000000(0) | 0.000000 (0) | 0.000000 (0) | 0.01526 | 0.375 |
|  | Mn | 0.000000 (0) | 0.000000 (0) | 0.000000 (0) | 0.02500 | 0.625 |
|  | O1 | 0.449329 (0) | 0.000000 (0) | 0.250000 (0) | 0.01675 | 1.0 |
| 0.75 | a (Å) = 5.51421 (0.00010) b (Å) = 5.51421 (0.00010) c (Å) = 13.37500 (0.00048) | | | | | |
|  | La | 0.000000 (0) | 0.000000 (0) | 0.250000 (0) | 0.03017 | 1.0 |
|  | Fe | 0.000000(0) | 0.000000 (0) | 0.000000 (0) | 0.01319 | 0.25 |
|  | Mn | 0.000000 (0) | 0.000000 (0) | 0.000000 (0) | 0.02500 | 0.75 |
|  | O1 | 0.451235 (0) | 0.000000 (0) | 0.250000 (0) | 0.00376 | 1.0 |
| 0.875 | a (Å) = 5.51379 (0.00000) b (Å) = 5.51379 (0.00000) c (Å) = 13.38999 (0.00000) | | | | | |
|  | La | 0.000000 (0) | 0.000000 (0) | 0.250000 (0) | 0.02500 | 1.0 |
|  | Fe | 0.000000(0) | 0.000000 (0) | 0.000000 (0) | 0.02500 | 0.125 |
|  | Mn | 0.000000 (0) | 0.000000 (0) | 0.000000 (0) | 0.02500 | 0.875 |
|  | O1 | 0.450197 (0) | 0.000000 (0) | 0.250000 (0) | 0.02500 | 1.0 |
| 1.0 | a (Å) = 5.51053 (0.00018) b (Å) = 5.51053 (0.00018) c (Å) = 13.39558 (0.00074) | | | | | |
|  | La | 0.000000 (0) | 0.000000 (0) | 0.250000 (0) | 0.04060 | 1.0 |
|  | Mn | 0.000000 (0) | 0.000000 (0) | 0.000000 (0) | 0.01873 | 1.0 |
|  | O1 | 0.465888 (0) | 0.000000 (0) | 0.250000 (0) | 0.03805 | 1.0 |